\documentclass[a4paper,amsmath,amssymb,superscriptaddress,aps,prb,reprint,showpacs]{revtex4-1}
\usepackage{color,graphicx,pstricks}
\usepackage{subfigure}
\usepackage{MnSymbol}

\newcommand{\cdag}{c^{\dagger}}
\newcommand{\cnod}{c^{\phantom{\dagger}}}
\newcommand{\e}{\textrm{e}}

\begin{document}
\title{From Frustrated to Unfrustrated: Coupling two
  triangular-lattice itinerant Quantum Magnets}

\author{Sahinur Reja}
\affiliation{Department of Physics, Indiana University, Bloomington, Indiana 47405, USA}
\affiliation{IFW Dresden, Helmholtzstr. 20, 01069 Dresden, Germany}
\author{Pavel S. Anisimov}
 \affiliation{Institute for Functional Matter and Quantum Technologies, University of Stuttgart,
Pfaffenwaldring 57
70550 Stuttgart, Germany}
\author{Maria Daghofer}
 \affiliation{Institute for Functional Matter and Quantum Technologies, University of Stuttgart,
Pfaffenwaldring 57
70550 Stuttgart, Germany}
\affiliation{Center for Integrated Quantum Science and Technology, University of Stuttgart,
Pfaffenwaldring 57
70550 Stuttgart, Germany}

\begin{abstract}
Motivated by systems that can be seen as composed of two frustrated
sublattices combined into a less frustrated total lattice, we study
the double-exchange model with nearest-neighbor (NN) and
next--nearest-neighbor (NNN) couplings on the honeycomb lattice. When adding
NN hopping and its resulting double exchange to the antiferromagnetic (AFM) Heisenberg coupling, the resulting phase diagram is
quite different from that of purely Heisenberg-like magnetic models and strongly depends on
electron filling. For half filling, patterns of AFM dimers dominate, where the
effective electronic bands remain graphene-like with Dirac cones in all phases, from the FM
to the $120^\circ$ limit. When the density of states at the Fermi level is
sizable, we find non-coplanar incommensurate states as
well as a small-vortex phase. Finally, a non-coplanar
commensurate pattern realizes a Chern insulator at
quarter filling.
In the case of both NN and NNN hopping, the
noncoplanar spin pattern inducing Chern insulators in triangular
lattices is found to be quite stable under coupling into a honeycomb system. The
resulting total phases are topologically nontrivial and either a Chern
insulator with $C=2$ or a magnetic topological crystalline insulator protected
by a combination or mirror-reflection and time-reversal symmetries arise. 
\end{abstract}

\date{\today} 


\maketitle

\section{Introduction}

Frustration in magnetic systems is currently a topic of high interest,
because it can support a variety of unconventional states by
suppressing more standard ordered states that may arise without it. Examples are
spin liquids~\cite{Fiete:ge,Balents:2010ds} and spin ice,\cite{Bramwell:2001ex,Huang:kq} where magnetic moments do not fully order
even at the lowest temperatures, or non-coplanar magnetic
patterns. The latter may arise may through an interplay of spin-orbit coupling
and magnetism, e.g. in the case of skyrmions~\cite{Rossler:2006cq} and vortex crystals,~\cite{PhysRevB.93.104417,PhysRevX.4.011023} or via the
competition~\cite{Martin08,Ueland:2012ju,2015arXiv151006830O} of antiferromagnetic (AFM) superexchange and kinetic
energy of electrons, which tends to prefer ferromagnetic (FM)
order. 

Frustration can have various sources, from lattice geometry over
orbital symmetries to more generally competing tendencies in a
Hamiltonian. Accordingly, it can also be lifted in a variety of
ways. Here we focus on frustration due to lattice geometry in combination with
competition between kinetic and super-exchange energy. Frustration
might then on one hand be lifted by changing the balance between kinetic
and magnetic terms (e.g. by doping), but on the other hand also by
changing the lattice from frustrated to unfrustrated. This second
mechanism is currently under debate in double-perovskite systems,
which are composed of two interpenetrating face-centered--cubic
lattices, made up from ions of one element each.~\cite{PhysRevB.63.180407,Morrow:2013ka,Meetei:2013fd,PhysRevLett.112.147202} Each separate lattice
is then highly frustrated, but the combined one is simple cubic and
unfrustrated. How strong the sublattices are connected -- and thus the
degree of frustration -- depends on the specific compound. 

Here, we address such a mechanism in a much simpler
setting, namely by coupling two triangular lattices (frustrated) into
a honeycomb one (unfrustrated). This geometry is potentially also
applicable to materials, namely to bilayers,~\cite{PhysRevLett.99.157203} where each layer forms a
triangular lattice and where the top layer is shifted with respect to
the bottom layer, so that each `top' site lies in the middle of a
triangle formed by `bottom' sites. With this in mind, the half-filled
Hubbard model has recently been studied for the same geometry, i.e., a
honeycomb model with (unfrustrated) first- and (frustrated)
second-neighbor hopping.~\cite{Jiang:2015dg} The phase diagram was
shown to contain stripes, spirals and phases with non-coplanar
magnetism with trivial and non-trivial Chern numbers $C=0,2$, which
were ascribed mainly to magnetic frustration of the super-exchange
mechanism. Interestingly, the spin pattern giving  $C=2$ in the  frustrated
half-filled honeycomb lattice has also been reported to be
stabilized by a van-Hove singularity in the unfrustrated
quarter-filled model without second-neighbor hopping.~\cite{Li:2012dt,PhysRevB.85.035414} 

We focus here on the Kondo-lattice model describing itinerant
electrons coupled to localized magnetic moments. On one hand, this is
motivated by the potentially more complex physics of double
perovskites with localized and itinerant carriers,~\cite{Paul:2013dy} on the other, magnetic interactions and electron
itineracy can here be
addressed  separately. The Kondo-lattice model has been studied
in detail on both the triangular and the honeycomb lattice. On the
triangular lattice, one of the most intriguing phases is a quantum
anomalous Hall (QAH) state, where a non-coplanar spin pattern induces
non-trivial band topology.\cite{Martin:2008dx,Kumar:2010p216405} At the same density, the honeycomb system
shows in addition to straightforward FM and N\'eel AFM states also
effects such as massive, but sub-extensive,  ground-state degeneracy and emergent
frustration due to a spontaneous triangular superstructure.\cite{Venderbos:2011KLM} 
Perhaps not surprisingly, we will find here that coupling two triangular
lattices yields phases reminiscent of either case. 

Without kinetic energy, i.e. for localized Heisenberg spins, the
interpolation between triangular and honeycomb lattices has been
addressed extensively. The resulting $J_1$-$J_2$(-$J_3$) honeycomb
model has revealed a rich phase diagram with several incommensurate,
but coplanar, 
spiral phases.\cite{Rastelli:1979ie,Fouet:2001kf} We are going to see here how the kinetic
energy of charge carriers modifies this picture and that it can 
not be captured by an effective FM coupling. One important difference is
that the kinetic energy will be seen to favor commensurate patterns or
non-coplanar states over incommensurate spirals. Finally, issues related to ours have also been discussed\cite{Li:2016tz}
for a frustrated honeycomb Kondo-Heisenberg model. In this study,
localized spins were in the quantum limit  $S=1/2$ and
antiferromagnetically coupled to electrons; an important result is the
replacement of N\'eel antiferromagnetism by disordered valence-bond
states. In our case of large spins, whose moment cannot be balanced by the
conduction electrons, order is not destroyed, even though we similarly
find phases, where electron motion is confined to small units like  dimers or 
hexagons. 

We are going to analyze two scenarios, a `purely magnetic' frustration,
where hopping is restricted to nearest neighbors and where 
nearest-neighbor (NN) and next--nearest-neighbor (NNN) spin exchange interpolates between the
limits of N\'eel-AFM order and the $120^\circ$ pattern of the
Yafet-Kittel state. Among the variety of phases supported by this
interplay, we find effects like vortex patterns and non-coplanar
spiral-like phases, but also a QAH phase with Chern number $C=1$.

The second scenario involves strong
second-neighbor hopping, where each triangular lattice is in a
topologically non-trivial QAH state. When coupling them
together into a honeycomb system, we can find
a Chern insulator with $C=2$, as the phase reported for the Hubbard
model.~\cite{Li:2012dt,PhysRevB.85.035414,Jiang:2015dg} There is additionally a phase with opposite
chiralities in the sublattices, which has Chern number $C=0$, but is
nevertheless topologically nontrivial: a combination of time reversal and mirror
reflection protects edge states on zig-zag edges, lading to a state
similar to a topological  crystalline insulators
(TCI),~\cite{Fu:2011ia} but with underlying magnetic order.

After introducing the model and the method in
Sec.~\ref{sec:model_method}, we present results with unfrustrated NN
hopping and purely magnetic frustration in Sec.~\ref{sec:purely_J}. We
discuss two electron densities, half filling $n=1/2$ (Sec.~\ref{sec:n_1_2})
and $n=2/3$ (Sec.~\ref{sec:n_2_3}) with very
different density of states at the Fermi level in order to draw out
the impact of the kinetic energy. We then
include second-neighbor hopping  in  Sec.~\ref{sec:kin_frustr}, where
we  focus on $n=1/2$, because this filling shows the QAH state on the
triangular lattice. We discuss in particular the emergent TCI and its
symmetries Sec.~\ref{sec:TCI}. Section~\ref{sec:conclusion} offers a summary and
conclusion. 

\section{Model and Method}\label{sec:model_method}

We consider a honeycomb double-exchange model with NN and NNN
terms. For dominant NNN parameters, this corresponds to two triangular
lattices that are connected 
via the NN terms. The model describes itinerant fermions
that interact with localized spins via Hund's rule
coupling. For localized spins large enough to be treated classically
and dominant Hund's rule coupling, one can focus the discussion
to itinerant fermions whose spin is parallel to the localized
spins.~\footnote{For classical
  spins, antiferromagnetic coupling would lead to an equivalent
  situation.} If we assume that the localized spins moreover interact
with each other, the model Hamiltonian becomes 
\begin{align}
H &= - t_1 \sum_{ \langle ij \rangle} \left ( \Omega({\bf S}_{i},{\bf S}_{j})
 c^{\dagger}_{i} c^{~}_{j} + H.c. \right )\nonumber\\
&\quad
- t_2 \sum_{ \llangle ij \rrangle} \left (\Omega({\bf S}_{i},{\bf S}_{j}) c^{\dagger}_{i} c^{~}_{j} + H.c. \right )\nonumber\\
&\quad
+ J_{1} \sum_{ \langle ij \rangle} {\bf S}_{i} \cdot {\bf S}_{j} 
+ J_{2} \sum_{ \llangle ij \rrangle} {\bf S}_{i} \cdot {\bf S}_{j} \label{eq:ham}
\end{align}
where $c^{}_{i}$ ($c^{\dagger}_{i}$) is the annihilation (creation) 
operator for an electron with spin parallel to the local magnetic moment ${\bf S}_i$. 
Angular brackets $\langle ij \rangle$ and $\llangle ij \rrangle$
denote NN and NNN pairs of sites on a honeycomb lattice. $J_1$ and
$J_2$ give  the strengths of Heisenberg intersite coupling. $t_1$ and
$t_2$ parameterize NN and NNN hopping, which is however decisively
modified by the local spin structure: As the spin of electron always
has to point along the local quantization axis $\parallel {\bf S}_i$,
the relative spin orientation enters via $\Omega({\bf S}_{i},{\bf S}_{j}) =
[\cos(\theta_i/2)\cos(\theta_j/2) + \sin(\theta_i/2)\sin(\theta_j/2)
e^{{\rm -i} (\phi_i - \phi_j)}]$, with  polar and azimuthal angles 
\{$\theta_i$,$\phi_i$,$\theta_j$,$\phi_j$\}.~\cite{PhysRevB.54.R6819,Dag01}

The model is investigated using a Markov-chain Monte Carlo (MCMC) method which combines
the classical Monte Carlo for spins with the diagonalization of the
fermion degrees of freedom.~\cite{Dag01} The solution of a fermionic
problem is required at each Monte Carlo update step in order to obtain the 
electronic contribution to the total energy of a given classical
configurations of localized spins. As this fermion problem is a
non-interacting one, however, it remains tractable for larger
clusters and as the spins are classical, no sign problem arises in the
Monte Carlo. The fermions only interact via their impact on the localized
spins. As we are here interested in the ground-state phase diagram
rather than in temperature-driven phenomena, we complement the MCMC
with a subsequent numerical optimization starting from the last MCMC
configuration. 

We additionally implemented an alternative algorithm based on a
Chebychev expansion of local Green's functions rather than full
diagonalization.~\cite{PhysRevLett.102.150604} For each local MCMC
update, only four Green's functions need to be evaluated, which can be
achieved by a recursive Chebychev scheme based on two initial
vectors. The scheme has a much better scaling with the number of sites
$N$, only  $\propto N$ rather than $\propto N^3$ as for full
diagonalization.\footnote{When using full diagonalization, we made use of the
fact that the low coordination number of the model without NNN hopping
permits the use of faster routines for band-diagonal matrices, if
sites on the 2D lattice are numbered suitably.} However, the
Chebychev-based algorithm did not extend attainable system sizes
significantly, because our most difficult cases, the incommensurate
noncoplanar states discussed in Sec.~\ref{sec:n_2_3}, require very low temperatures.
In this regime, the number of Chebychev polynomials needed becomes so
large that runtimes would only be improved for cluster sizes beyond
the reach of full MCMC simulations.

We calculate the ground state energy using this Monte-Carlo method
at different values of the Hamiltonian's parameters. We then identify the respective
magnetic order and, where appropriate, compare the energy obtained in Monte Carlo with
that of the perfectly ordered state. In addition to looking at and
comparing real-space spin configurations, we make use of  the  spin structure
factor
\begin{align}\label{eq:ssf}
S({\bf q}) &= \frac{1}{N^2} \sum_{ij} {\bf S}_i \cdot {\bf
  S}_j \  e^{-{\rm i}{\bf q}\cdot({\bf r}_i - {\bf r}_j)}\;.
\end{align}
where $N$ is the number of sites and 
${\bf r}_i$, ${\bf r}_j$ are the positions of sites
$i$,$j$. As the honeycomb lattice has a two-site unit cell, momentum
${\bf q}$ from the first two Brillouin zones can be relevant. The
observable was evaluated from the optimized ground-state
configuration.

\section{Results and Discussion} \label{sec:results}

\subsection{Connecting two frustrated antiferromagnets}\label{sec:purely_J}

As a first example, we discuss NN hopping $t_1=1$ and NNN magnetic interaction
$J_2>0$, i.e., $t_2=0$ and $J_1=0$. This case thus describes two triangular Heisenberg
antiferromagnets that are coupled by electron itineracy. The kinetic
energy tends to favor FM spin alignment, because this maximizes band
width; the two limiting cases are thus (i) two decoupled triangular
lattices with $120^\circ$ pattern (for $J_2>0$ and $t_1=0$) and (ii)
one FM honeycomb lattice (for $J_2=0$ and $t_1\neq 0$).  Intuitively,
the intermediate case might then seem similar to the $J_1$-$J_2$ model
discussed in Refs.~\onlinecite{Rastelli:1979ie,Fouet:2001kf}, where 
coplanar incommensurate spirals interpolate between these limits. 

However, the rule of thumb that electron itineracy favors FM alignment
is often modified by more complex effects due to the phases of the
effective hoppings.~\cite{PhysRevB.54.R6819}  In particular, magnetic
patterns are favored if they open a band gap at the Fermi
level. The impact of the kinetic energy accordingly depends on the
band structure as well as on filling. We are here going to discuss two
fillings, $n=1/2$ (i.e. one electron per two
sites) and $n=1/3$ or $2/3$  (1 or 2 electrons per 3 sites). As the
kinetic energy is in this section restricted to
the NN bonds, i.e. lives on the bipartite honeycomb lattice, it is
particle-hole symmetric and $n=1/3$ is equivalent to $n=2/3$ in 
the limit of infinite Hund's rule studied here.    

For spinless fermions not coupled to any localizes spins, or alternatively
for an FM system, the two chosen filling fractions correspond to very different Fermi
surfaces: At $n=1/2$, the density of states vanishes in a pseudogap at the Fermi
level, while $n=1/3$, resp. $n=2/3$ corresponds to a substantial
density of states.  Mechanisms related to the kinetic energy can thus be
expected to play out quite differently in these two scenarios. One
feature where coupling via the kinetic energy has previously been
found to differ qualitatively from FM exchange is that all states
seen in the $J_1$-$J_2$(-$J_3$) model are coplanar, while
Kondo-lattice models are known to support non-coplanar states like the
(gapped) QAH state of the half-filled triangular
lattice.~\cite{Martin:2008dx}. Indeed, we are here going to see that
the phase diagram for $n=1/2$, with a weaker impact of the band
structure, consists predominantly of coplanar patterns while the
$n=1/3$/$n=2/3$ filling supports non-coplanar states over wide
parameter ranges.

\subsubsection{Dominant AFM dimers and coplanar order at $n=1/2$}\label{sec:n_1_2}

 \begin{figure}
\subfigure[]{ \includegraphics[width =\columnwidth,angle=0]{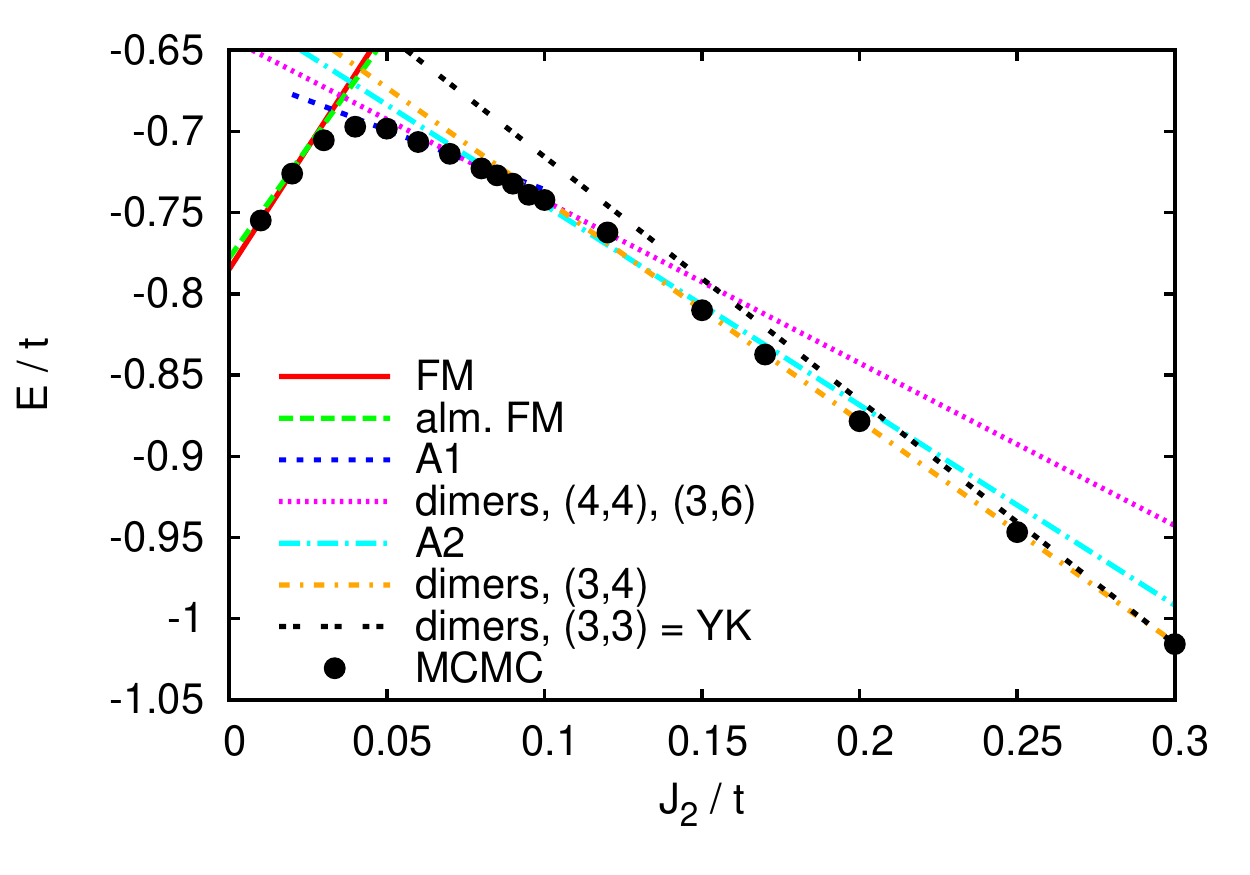} \label{phase_diag12}}
\subfigure[]{ \includegraphics[width =\columnwidth,angle=0]{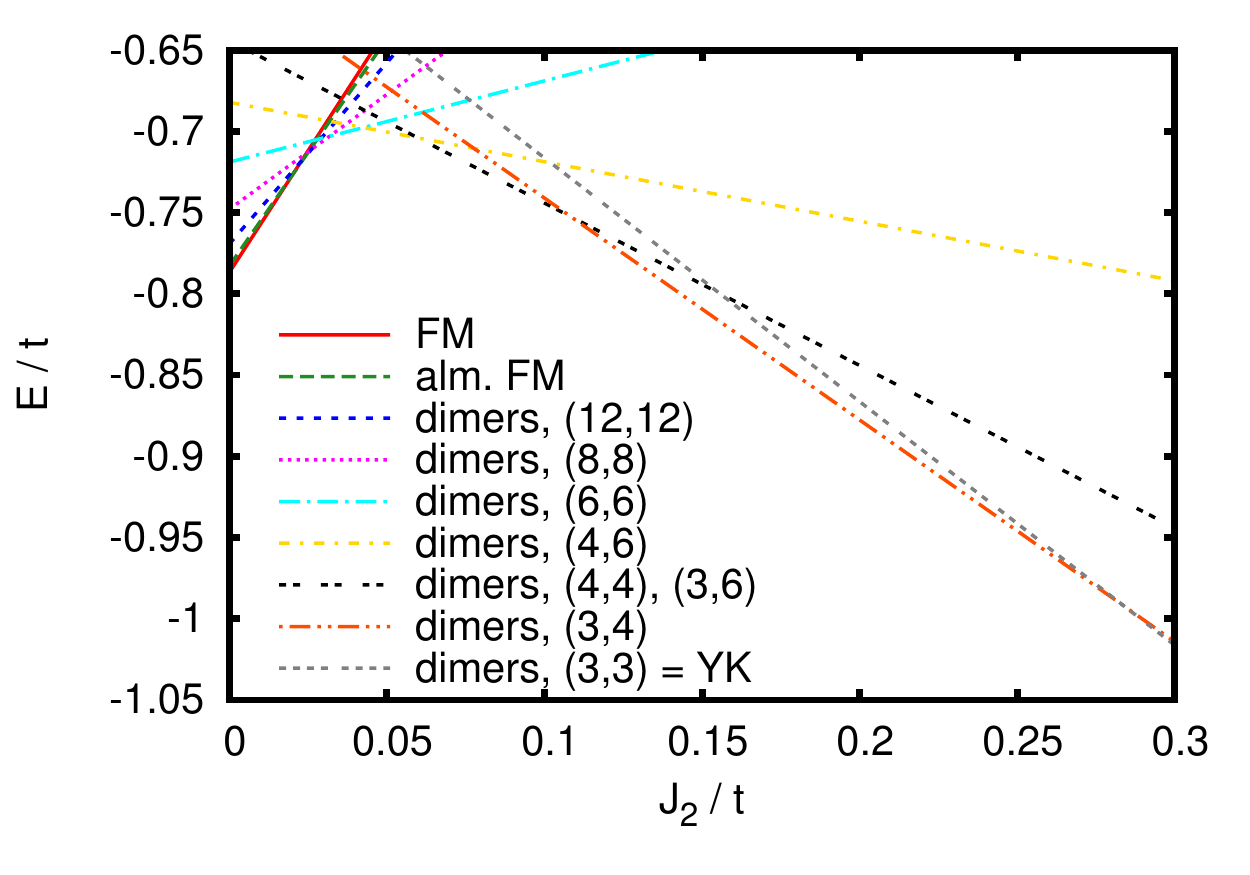} \label{phase_diag24}}
 \caption{(Color online) Different phases with increasing $J_2$ at
   $n=1/2$.  (a) Comparison MCMC energies (full circles) with the energies (straight
 lines) calculated for perfect spin configurations, lattice size
 $12\times 12$. (b) Ground-state energies of candidate phases inferred
 from (a), but for $24\times 24$ sites. Apart from the FM
 regime at very small $J_2$, most of the phase
 diagram is dominated by `dimer' phases, where FM dimers are arranged
 along one zig-zag direction. The phases can then be characterized by
 their periodicity along the other two -- still equivalent -- zig-zag
 directions. For the smallest unit cell with a periodicity of 3 along
 both axes, spins within each triangular sublattice form the
 $120^\circ$ pattern of the Yafet-Kittel (YK) state. $J_1=0$ and $t_2=0$\label{phase_diag_half}} 
 \end{figure}

 \begin{figure}
   \subfigure{\includegraphics[width =0.47\columnwidth,trim =0 60 40 50 , clip]{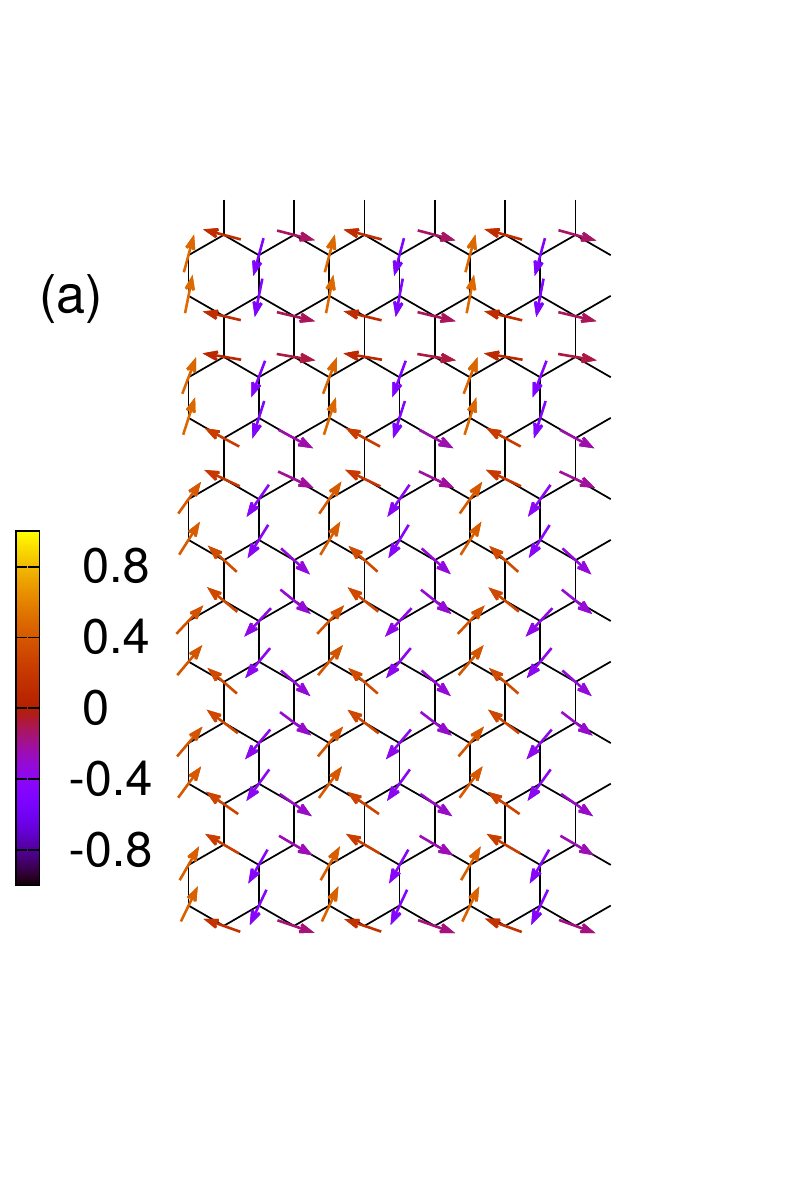}\label{fig:snap_08_12}}
   \subfigure{\includegraphics[width =0.47\columnwidth,trim =0 60 40 50 , clip]{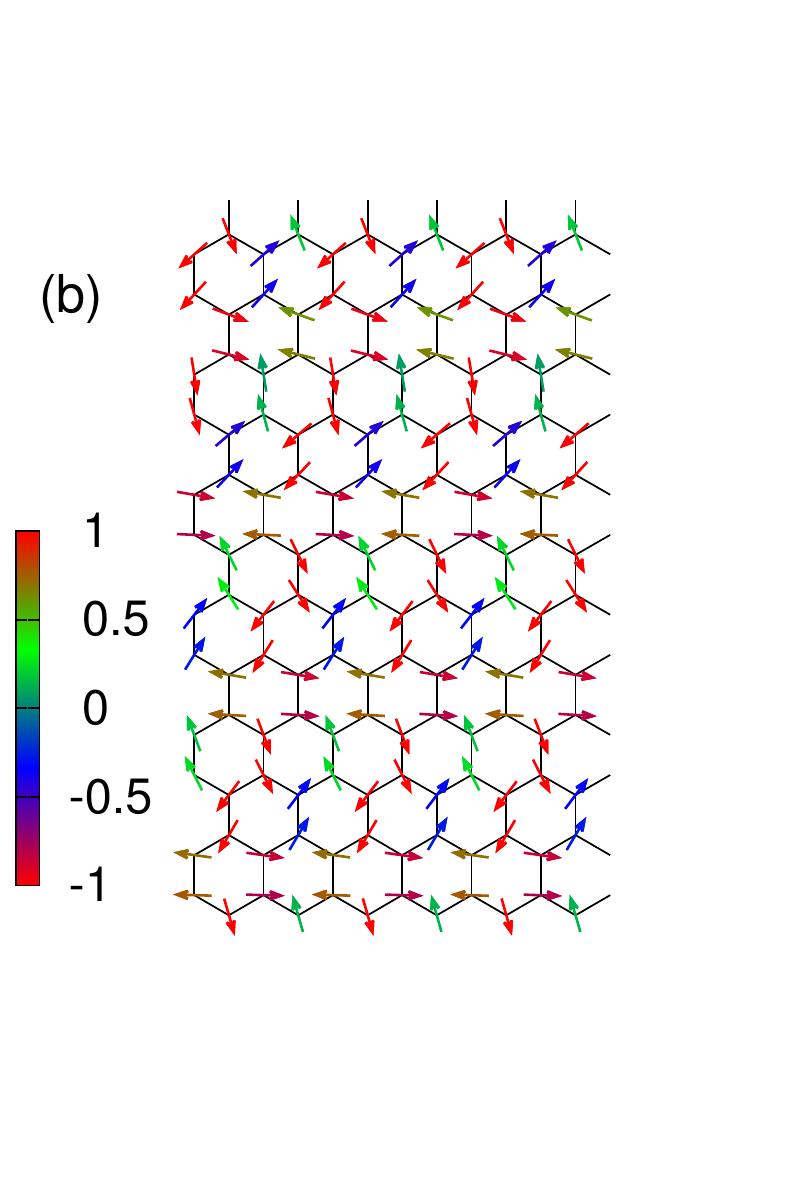}\label{fig:snap_1_12}}
   \subfigure{\includegraphics[width =0.47\columnwidth,trim =0 60 40 50 , clip]{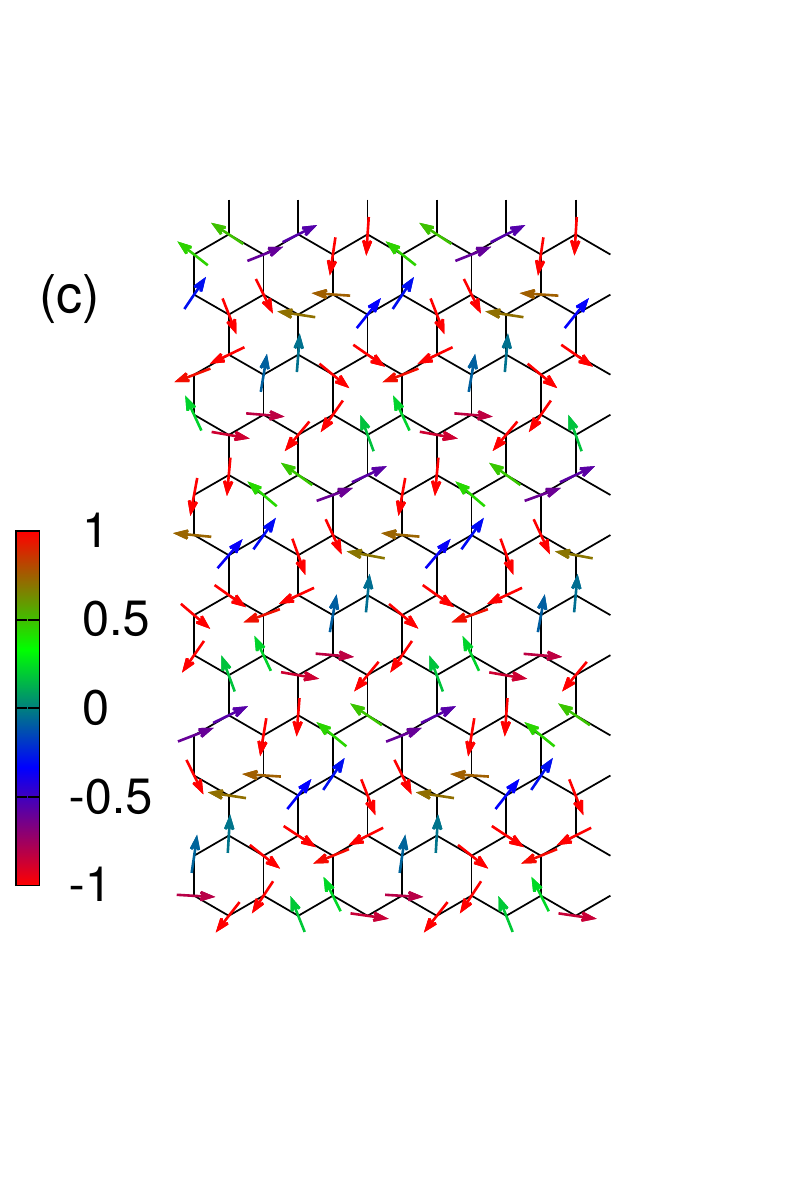}\label{fig:snap_2_12}}
   \subfigure{\includegraphics[width =0.47\columnwidth,trim =0 60 40 50 , clip]{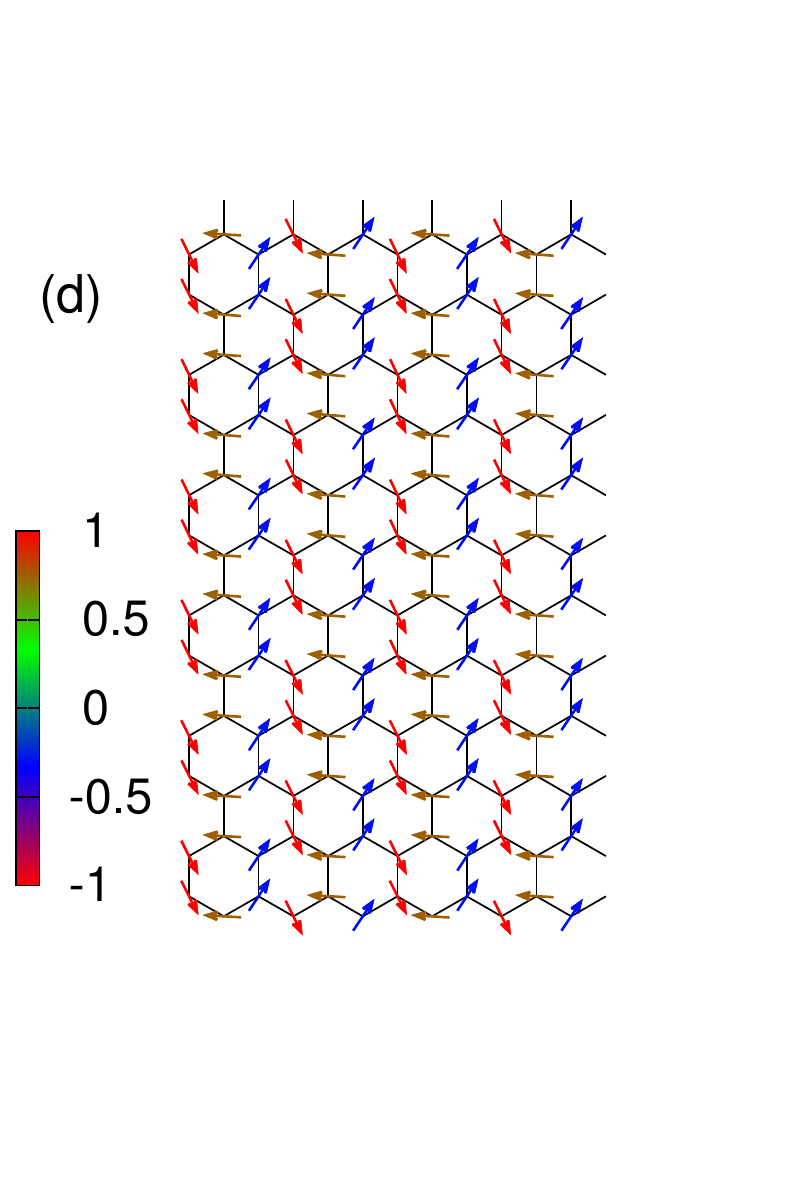}\label{fig:snap_3_12}}
   \caption{(Color online) Dimer phases with different periodicities as
     obtained by MCMC and optimization for $12\times 12$ sites. (a)
     Periodicities $(3,6)$ for $J_2 = 0.08 t_1$, (b)
     periodicities $(4,4)$ for $J_2 = 0.1 t_1$,  (c)
     periodicities $(3,4)$ for $J_2 = 0.2 t_1$,  (d)
     periodicities $(3,3)$ for $J_2 = 0.3 t_1$. The last pattern has the
     $120^\circ$ YK pattern within each sublattice.\label{dimer_phases}}
 \end{figure}

Without $J_2$, the system is of course FM. MCMC simulations for
$12\times 12$ sites at small $J_2$ show a spiral that is FM along an
armchair edge and has the largest period allowed by the lattice along
zig-zag. This effect is well known for KLMs;~\cite{PhysRevLett.79.713,0953-8984-9-19-002,0953-8984-9-11-001,Venderbos:2011KLM}  as the spiral winning
over FM order always has the largest possible period and thus
converges to the FM state in the thermodynamic limit, it should be
considered a finite-size effect. It shows, however, that our simulations
reliably resolved the small energy difference to the FM pattern. 

For vanishing $t_1=0$, we expect each sublattice to show the
$120^\circ$ pattern of the Yaffet-Kittel state, and this is indeed
found for $J_2 \gtrsim 0.3 t_1$. $J_2$ can by itself of course
not decide the relative orientation of the two sublattices. Moreover, $J_1$  cannot fix the
relative orientation either, because the
three nearest neighbors of each spin are always oriented along the
three directions of the $120^\circ$ pattern, so that the total NN
interactions always cancel. In the $J_1$-$J_2$ Heisenberg model,
incommensurate order, which distorts the $120^\circ$ arrangement,
arises instead. Here, however, finite $t_1$ stabilizes a pattern with
perfect $120^\circ$ order within the sublattices and where each spin  
has one FM nearest neighbor; the NN FM dimers lie along one of
the lattice's zig-zag directions, see Fig.~\ref{fig:snap_3_12}.

Such FM dimers turn out to dominate most of the phase diagram. The original
three-fold rotation symmetry is broken by the dimers: they are put onto the
lattice so that exactly one of the zig-zag directions does \emph{not} contain
dimers. The relative orientation of the dimers can then have different
periodicity along the two remaining equivalent zig-zag directions. The
$120^\circ$ Yaffet-Kittel pattern found at large $J_2$ corresponds to periods
of three along both directions. Intermediate values of $J_2$ stabilize
larger--unit-cell patterns, on the $12\times 12$ lattice, we find, e.g.,
$(6,3)$ and $(4,4)$ (with the same energies) and $(4,3)$ at rather large
$J_2$, see  Fig.~\ref{dimer_phases}. 

The dimer phases compete with spiral phases that can be seen as modifications
of an '$A$-phase' pattern where each spin has two FM and one AFM neighbor. In
this underlying pattern, the
FM bonds then run along a zig-zag direction, orientation between zig-zag
chains is AFM. The patterns actually seen in the simulations are modifications
where either the FM zig-zag ('$A2$') or the AFM armchair ('$A1$') direction has an additional
modulation with periodicity 12, i.e., the largest one compatible with the
lattice. In the latter $A1$ case, the FM direction is moreover not perfectly
FM, but involves some canting; this canting angle was numerically optimized
for the energy comparison to the MCMC data shown in
Fig.~\ref{phase_diag12}. Finally, MCMC results at the transition from the
'almost FM' to the '$A1$' phase ($J_2 \approx 0.04$) are unclear: there might
be a small parameter window with non-coplanar order.

While MCMC simulations are not feasible for much larger lattices, comparison
of ground-state energies was extended to a $24 \times 24$-site lattice, see
Fig.~\ref{phase_diag24}. We compared all phases seen on the smaller lattice,
with the large periodicity of the 'almost FM' and 'almost $A$ phase' spirals
adjusted to the larger lattice, as well as all dimer patterns compatible with
$24\times 24$ sites. Of these, the Figure only includes phases that are the ground state
for some range of $J_2$. We see that the  modulated $A$ phases are now
replaced by dimer arrangements, whose unit cell shrinks with increasing
$J_2$. 

At half filling, the competition between NNN $J_2$ and NN $t_1$ thus largely
plays out via the formation of FM dimers, which are themselves arranged in
periodic patterns. For rather  small $J_2 \lesssim 0.1$ close the FM phase,
the periodicity becomes large, so that the thermodynamic limit can show
incommensurate order. For intermediate and large $J_2$, patterns
are commensurate, in contrast to the $J_1$-$J_2$ Heisenberg model. Similar to
Heisenberg models, on the other hand, is the fact that observed phases are
coplanar.  

Interestingly,
the one-particle spectral density for all dimer patterns remains  -- as for the FM
phase --  graphene-like with Dirac cones. Hoppings are
anisotropic, as they are reduced for two of the three honeycomb bond
directions: hoppings remain $t_1$ for the bonds supporting dimers,
they are $t_1\cos (\pi/n_{a/b})$, with $n_{a/b}$ referring to the two
periodicities, along the other two kinds of bonds. Accordingly, band
width is somewhat reduced and the location of the Dirac cones
changes, but features like the linear dispersion and the pseudogap
remain. While the electronic energy contribution is modified by the
renormalized band width, dimer states are not stabilized by the
opening of a band gap -- at most, they can profit from the fact that the
Dirac-cone pseudogap is not filled.

The pure honeycomb double-exchange model at half filling, i.e.
for $J_2=0$, has been studied in Ref.~\onlinecite{Venderbos:2011KLM}
and the dimer motif discussed here connects to this previous case. For
relatively strong $J_1 \geq 0.2 t_1$, the $J_1$-$t_1$ models is dominated by
FM or slightly canted dimers that lie on bonds along one zig-zag direction and
AFM coupled. The salient feature is that after fixing directions on one
zig-zag line, two degenerate possibilities exist for assigning spins on each
subsequent line. As a result, the system has a large, but subextensive,
degeneracy. In contrast to similar phases in compass models,\cite{RevModPhys.87.1} but
similar to an even more complex model involving additionally lattice degrees of
freedom,~\cite{Liang:2011fe} this is not
connected to any underlying symmetry of the Hamiltonian. We verified that this
physics remains stable in the presence of $J_2$, e.g. for $J_1=0.5 t_1$ and 
$J_2=0.2 t_1$. This corroborates the previous finding that such highly
degenerate phases building on emergent symmetries arise readily in
double-exchange models.

\subsubsection{Incommensurate and non-coplanar states at $n=2/3$}\label{sec:n_2_3}

Let us now discuss a filling of $n=2/3$, where the Fermi level for the
FM system falls into a region with quite a high density of states. Accordingly, the kinetic energy can be expected to have a
larger impact. We indeed consistently find one signature feature that
distinguishes frustrated itinerant-electron physics from Heisenberg spin models:
for intermediate $J_2,$, the phase diagram is dominated by non-coplanar states.

\begin{figure}
\subfigure[]{\includegraphics[width = 0.47\columnwidth,  trim=30 0 50 0,clip]{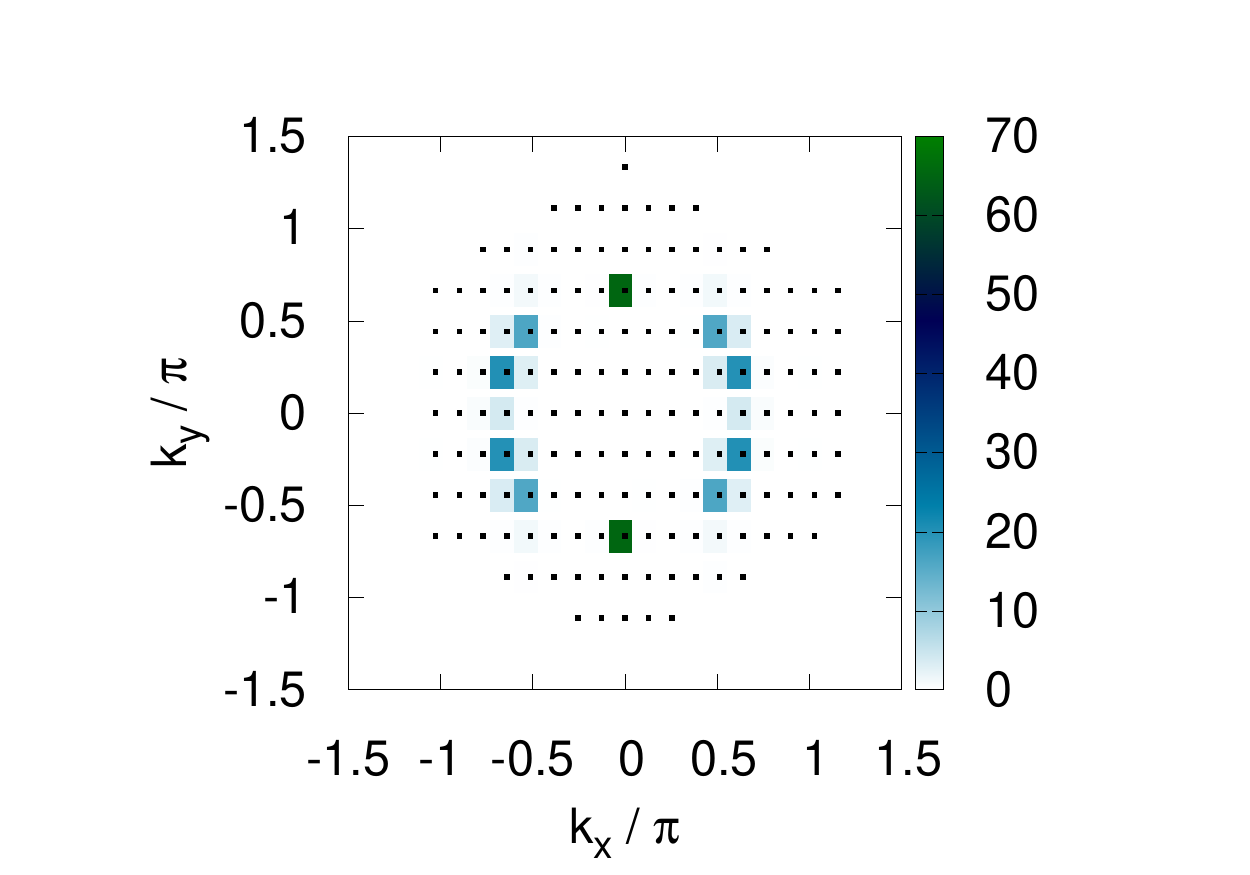}}
\subfigure[]{\includegraphics[width = 0.47\columnwidth,  trim=30 0 50 0,clip]{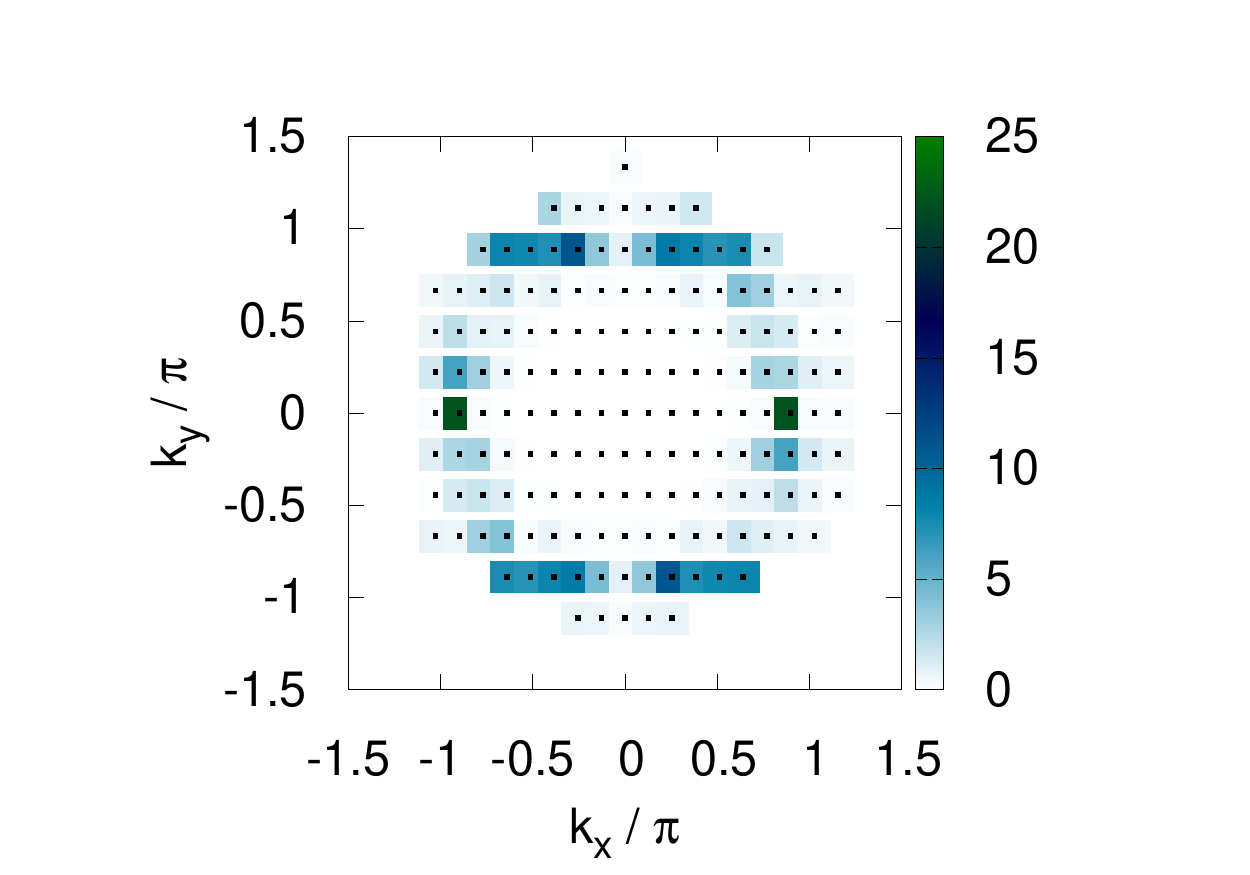}}\\[-0.5em]
\subfigure[]{\includegraphics[width = 0.47\columnwidth,  trim=30 0 50 0,clip]{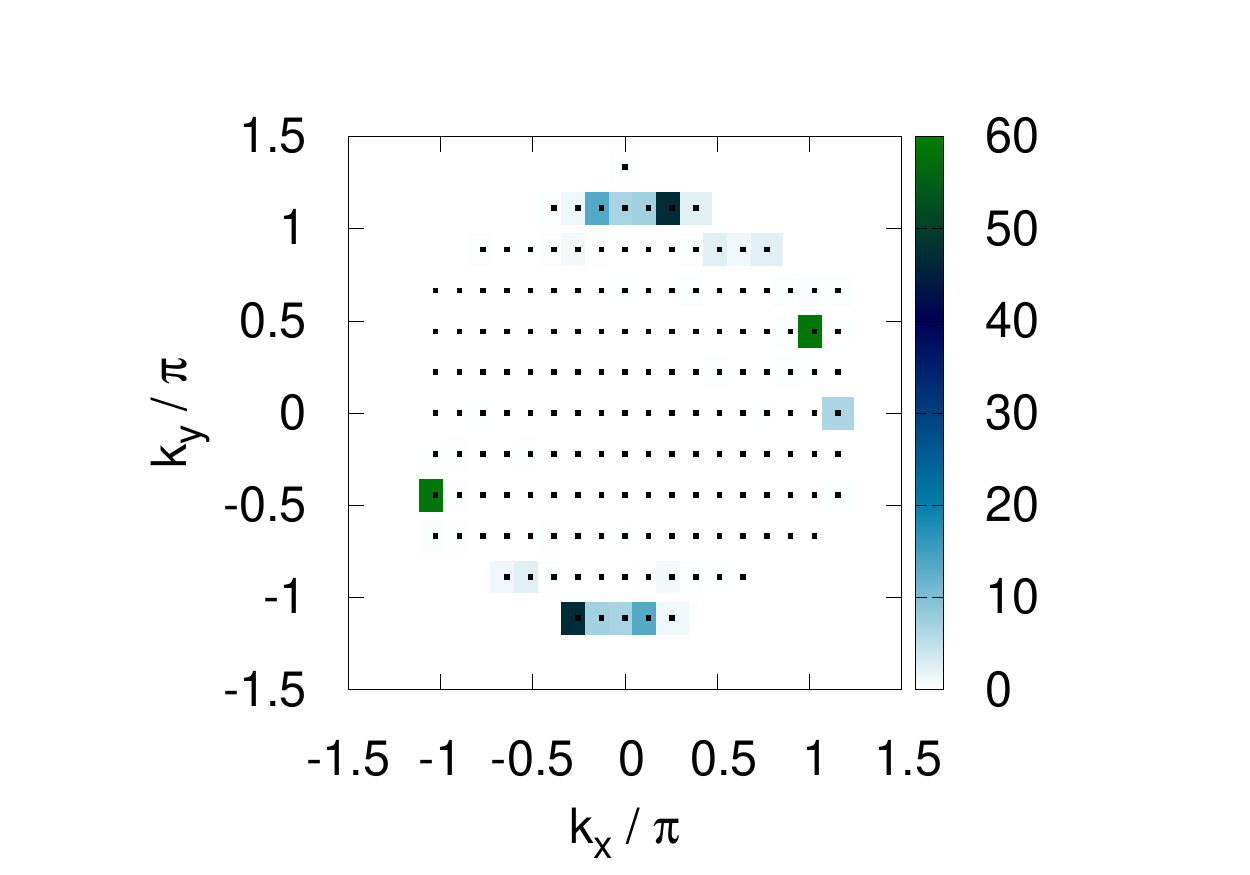}}
\subfigure[]{\includegraphics[width = 0.47\columnwidth,  trim=30 0 50 0,clip]{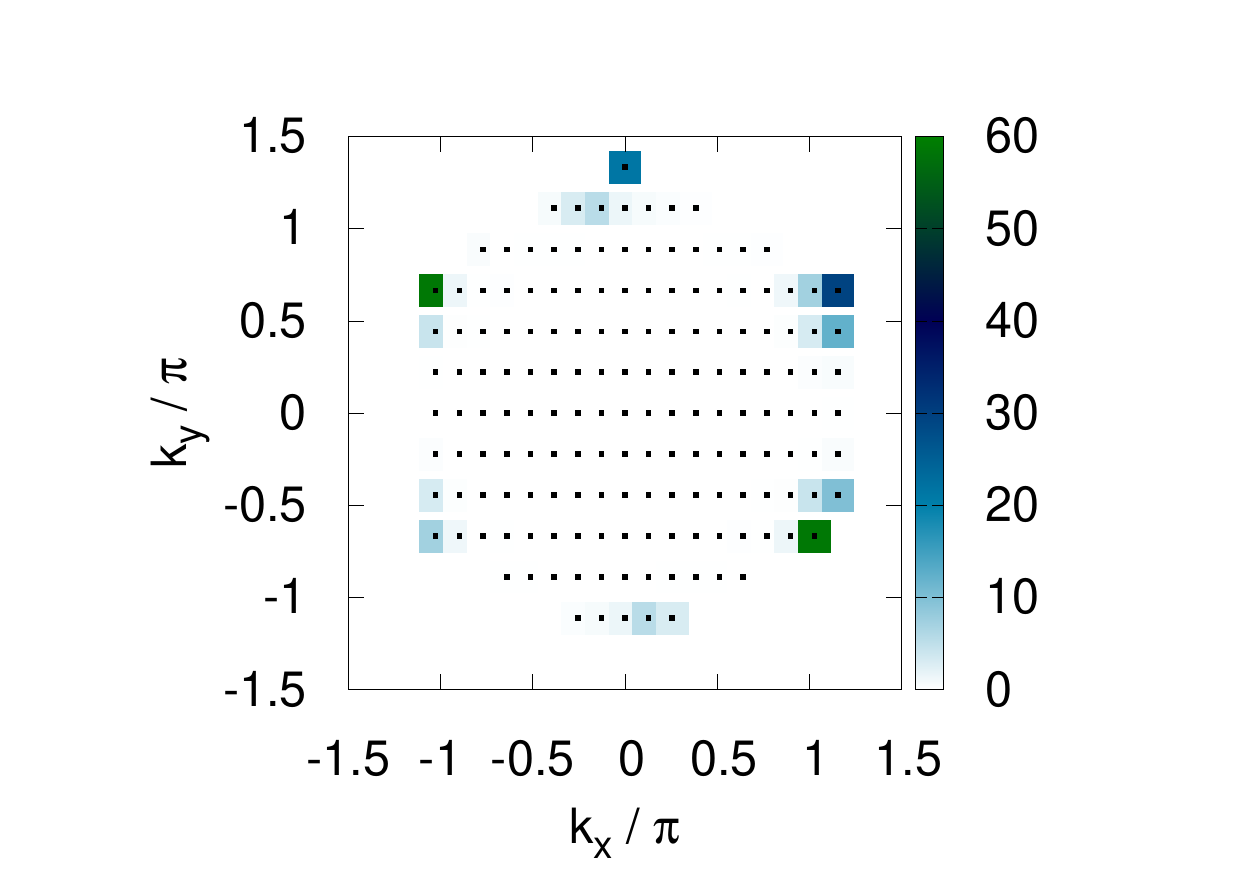}}\\[-0.5em]
\caption{Spin-structure factor in the first Brillouin zone for a filling of
  $n=2/3$ (or $n=1/3$) and (a) $J_2 = 0.02 t_1$, (b) $J_2 = 0.04 t_1$, (c)
  $J_2 = 0.12 t_1$ and (d) $J_2 = 0.16 t_1$. Obtained by MCMC and subsequent
  optimization for $18\times 18$ sites; $J_1=0$ and $t_2=0$.\label{fig:sk_n_2_3_j2}}
\end{figure}

\begin{figure}
\subfigure{\includegraphics[width =0.47\columnwidth,trim =0 60 40 50 , clip]{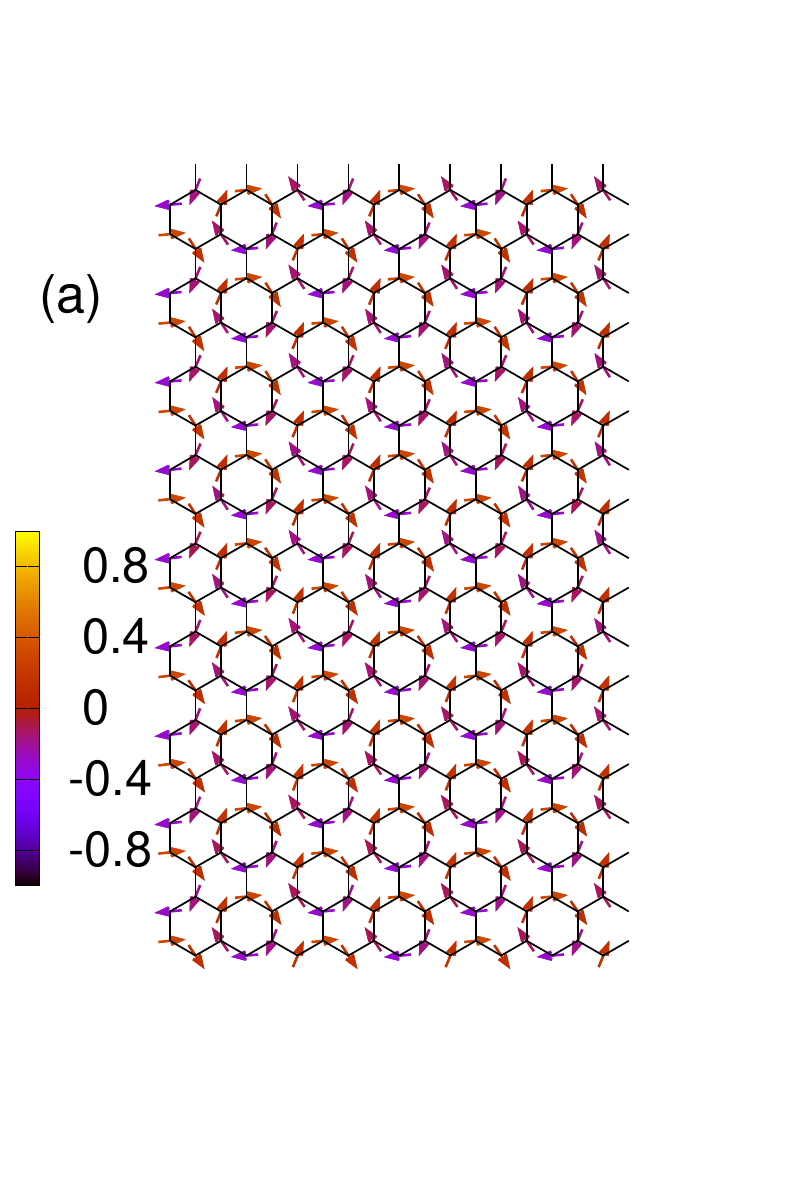}\label{fig:snap_3}}
\subfigure{\includegraphics[width =0.47\columnwidth,trim =0 60 40 50 , clip]{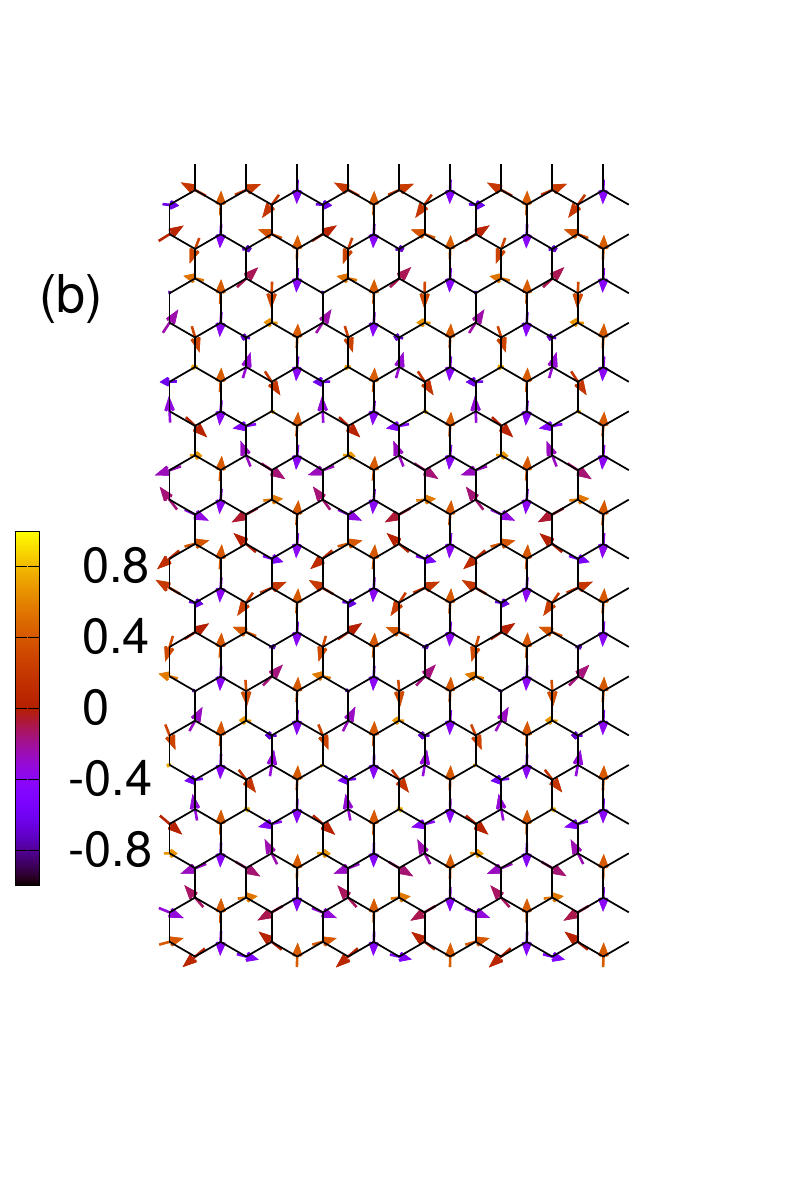}\label{fig:snap_25}}
\subfigure{\includegraphics[width =0.47\columnwidth,trim =45 0 60 0 , clip]{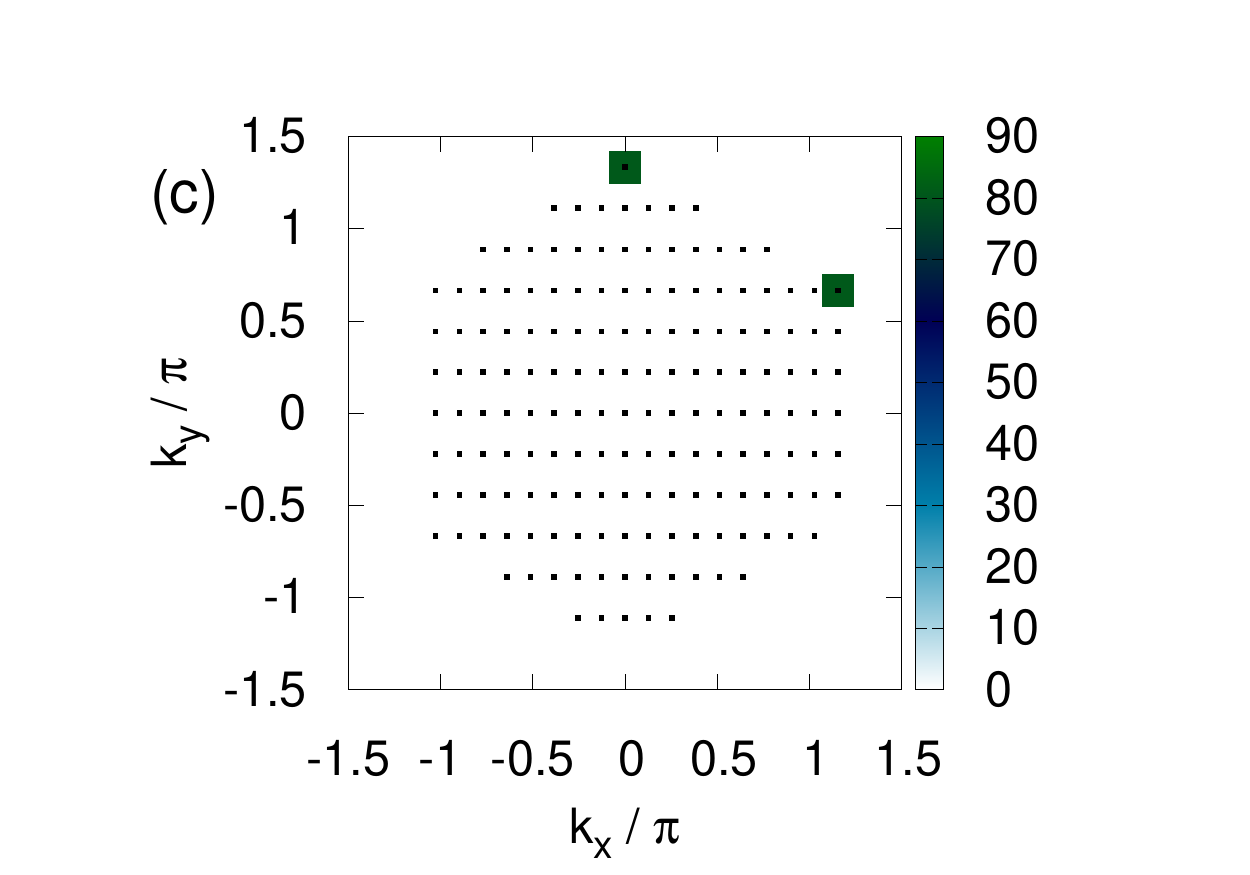}\label{fig:sk_3}}
\subfigure{\includegraphics[width =0.47\columnwidth,trim =45 0 60 0 , clip]{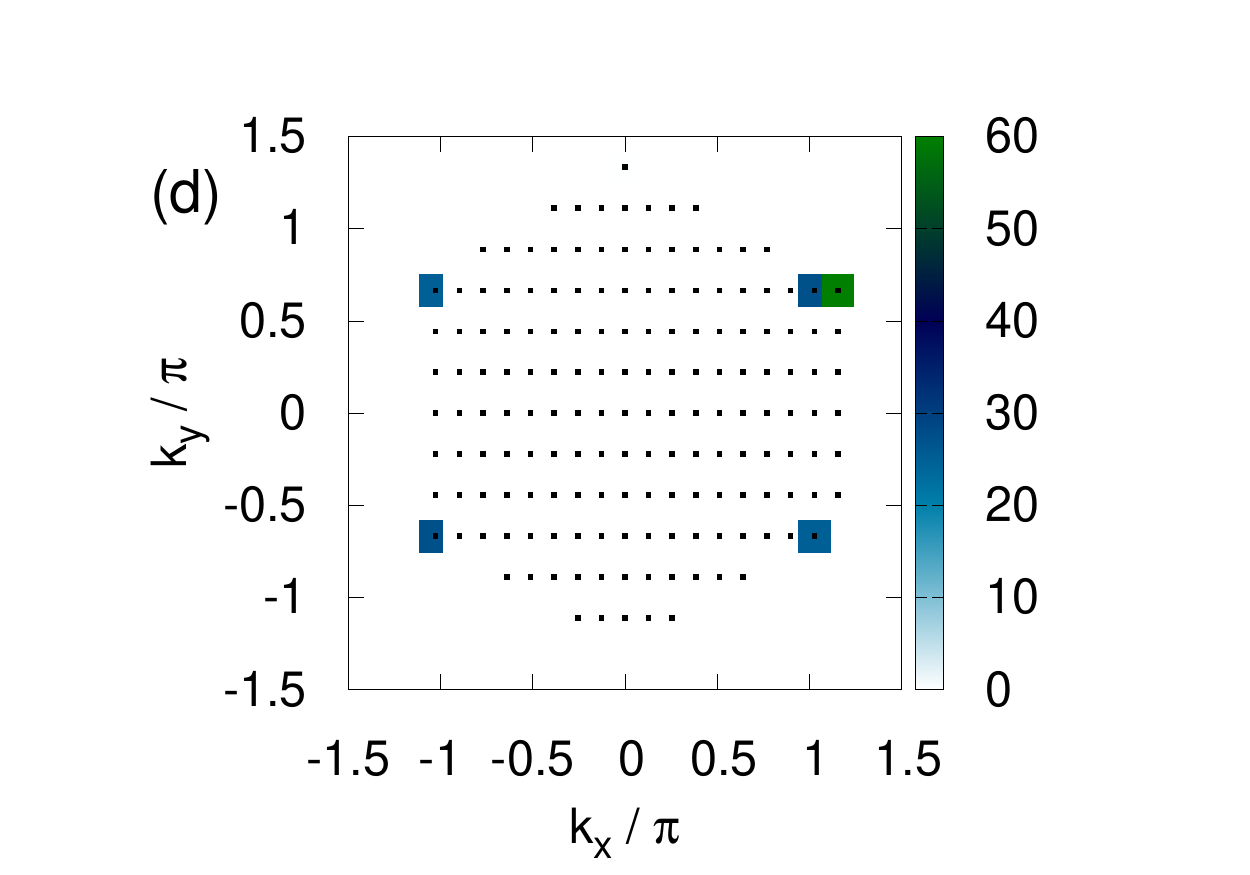}\label{fig:sk_25}}
\caption{(Color online) Disconnected vortices for large $J_2$. 
and a filling of  $n=2/3$ (equivalent to $n=1/3$). The disconnected vortices
can be seen in (a), which was obtained by MCMC and additional numerical optimization for $18 \times 18 $
sites and $J=0.3 t_1$. (b) shows the slightly distorted (noncoplanar) pattern
found for $J_2=0.25 t_1$. (c) and (d) are the corresponding spin-structure
factors for $J=2=0.3 t_1$ and $J_2=0.25 t_1$, resp. 
$t_2=0$ and 
$J_1=0$.\label{fig:incomm_noncopl_3}
}
\end{figure}

Due to the frustration, these non-coplanar states are incommensurate
-- unfortunately, this makes finite-size effects much more severe. The
diagonalization of the independent-electron problem does not scale
well with system size; we nevertheless simulated the system for up to
$18\times 18$ sites. Many features remain qualitatively similar for
the whole range $0.02 \lesssim J_2 \lesssim 0.16$:  Spins are non-coplanar and several momenta, whose
distance from the zone center grows with $J_2$,  contribute
to the spin-structure factor, see
Fig.~\ref{fig:sk_n_2_3_j2}. Furthermore, the one-particle
spectral density (not shown) is at least suppressed, if not clearly gapped, at
the Fermi level.

In some of the observed patterns, spins point with a similar
probability in any directions, while they are not 
distributed so isotropically in others. Similarly, rotational symmetry is found to be broken for
many values of $J_2$, as can be seen in the spin-structure factor
shown in Fig.~\ref{fig:sk_n_2_3_j2}, but not always. Given the current limitations on system size, we cannot
establish whether this is a true phase competition or a finite-size
effect. There are no indications for appreciable charge-density modulations, nor
can we observe skyrmion-like patterns, as have recently been reported
for the triangular-lattice model.~\cite{PhysRevLett.118.147205} Even
though skyrmions would go together
with noncoplanar and incommensurate patterns, their identification and
clear separation from competeing phases might also  need
larger lattices.  We implemented an alternative algorithm based on local Green's functions rather than full
diagonalization,~\cite{PhysRevLett.102.150604} but this did not extend
available cluster sizes sufficiently. 

For large $J_2\gtrsim 0.3 t_1$, MCMC simulations for $18\times 18$
sites show that spins within each triangular layer form a
$120^\circ$ YK 
state in order to satisfy $J_2$, to that the spin-structure factor is
peaked at the $K$ point, see Fig.~\ref{fig:sk_3}.  At $n=2/3$,
the electronic kinetic energy aligns the two $120^\circ$ patterns so
that  a `vortex crystal' arises, where the vortices  form a triangular lattice. Going around a vortex, spins
rotate by $60^\circ$ between sites, see Fig.~\ref{fig:snap_3}. Each vortex is connected to  6 neighboring
vortices by NN bonds, but these bonds always contain spins with
perfectly AFM orientation. Electrons can thus not hop from vortex to
vortex, hence the name  `disconnected vortex' (DV) phase. 

The purely magnetic energy, i.e., the contribution from
$J_2$ (and actually even from any $J_1$) is the same as for the 
$120^\circ$ dimer pattern seen at $n=1/2$. The kinetic energy,
however, is very different from the graphene-like bands of the
dimer state. As vortices are connected by
AFM bonds, electrons are filled into levels
available on each six-site ring. States located on different vortices
are perfectly degenerate. At first sight, six sites with
periodic boundary conditions would lead us to expect four states,
corresponding to $k=0$, $\tfrac{\pi}{3}$, $\tfrac{2\pi}{3}$, and $\pi$, of
which the middle two would be twofold degenerate. However, the spin
canting within the ring not only reduces the hopping strength, but
also adds an additional phase factor so that the allowed momenta
become $k=\tfrac{\pi}{6}$,$\tfrac{\pi}{2}$, and $\tfrac{5\pi}{6}$, all
doubly degenerate.
This
phase is thus expected to be favorable at fillings like $n=1/3$ and
$n=2/3$, but not at $n=1/2$, where the Fermi level would be found at a
particularly \emph{high} density of states. 

For slightly smaller values of $J_2 \approx 0.25 t_1$, i.e., just
before the onset of the commensurate disconnected-vortex phase, we
find a very similar pattern that carries an additional incommensurate
and non-coplanar modulation on
long length scales, see the configuration in Fig.~\ref{fig:snap_25}. The spin-structure factor accordingly has
components slightly off the $K$ point in addition to a signal at $K$,
see Fig.~\ref{fig:sk_25}. The three sharp peaks of the density of
states 
split, 
but the splitting is almost symmetric and the Fermi level
remains within a gap, so that the change in kinetic energy is
small. Comparing $12\times 12$ and $18\times 18$ sites and noting that
the additional spin-structure peaks are as close to $K$ as possible,
we conclude that larger lattices would only show the
disconnected-vortex state at even higher values of $J_2$. However,
similar to the `almost FM' state discussed above, the pattern becomes
ever more similar to the DV state and the splitting in the density of states
ever smaller.

\subsection{Topologically non-trivial states related to non-coplanar Chern Insulators}\label{sec:kin_frustr}

We next address the states with topologically
nontrivial bands, which we find if either NN Heisenberg
exchange $J_1$ or NNN hopping $t_2$ are added into the mix.

\subsubsection{Chern Insulator at Quarter filling}

\begin{figure}
\subfigure[]{\includegraphics[width=0.6\columnwidth,trim=50 30 50 50,clip]{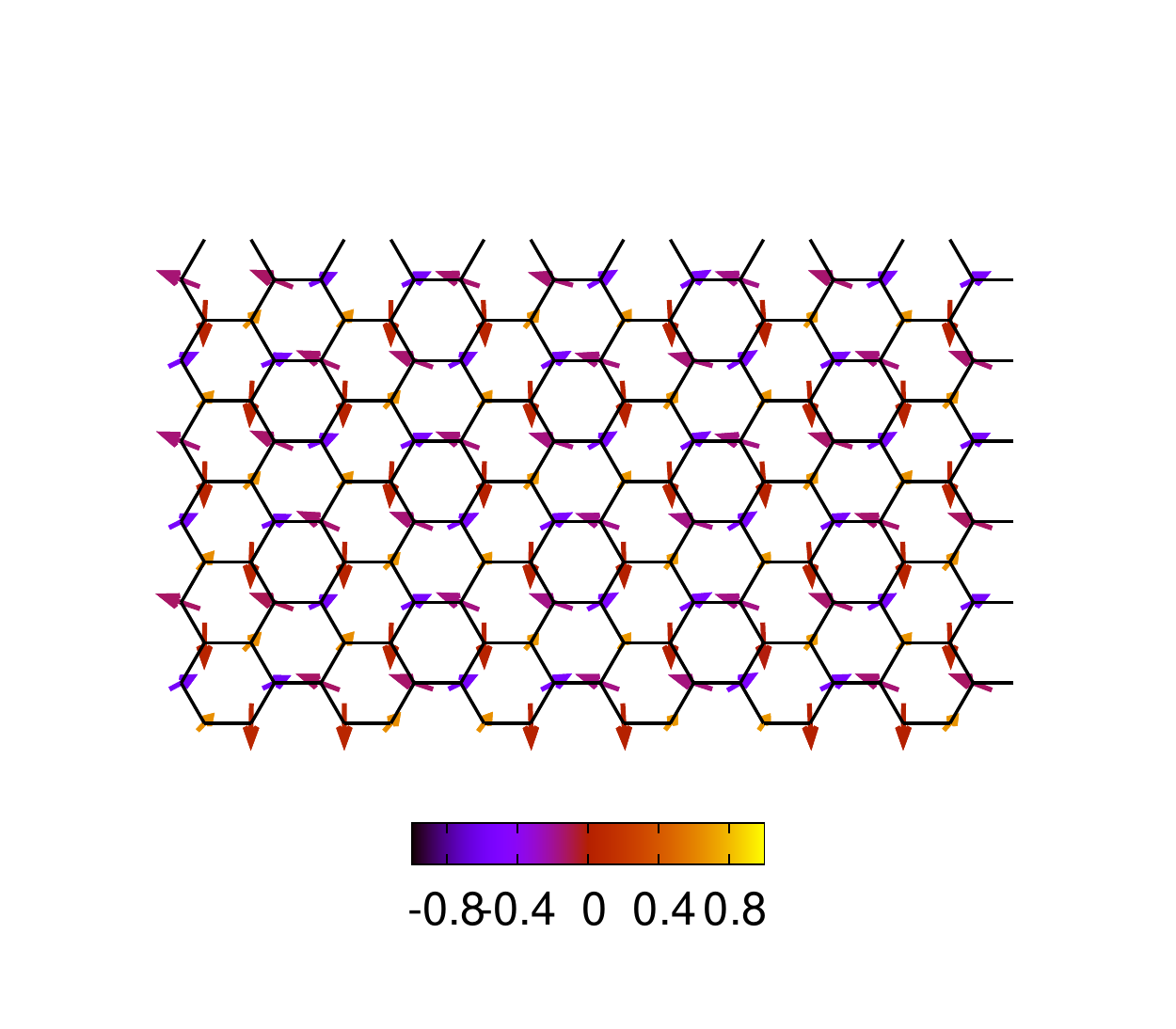}}
\subfigure[]{\includegraphics[width=0.37\columnwidth,trim=100 30 100
80,clip]{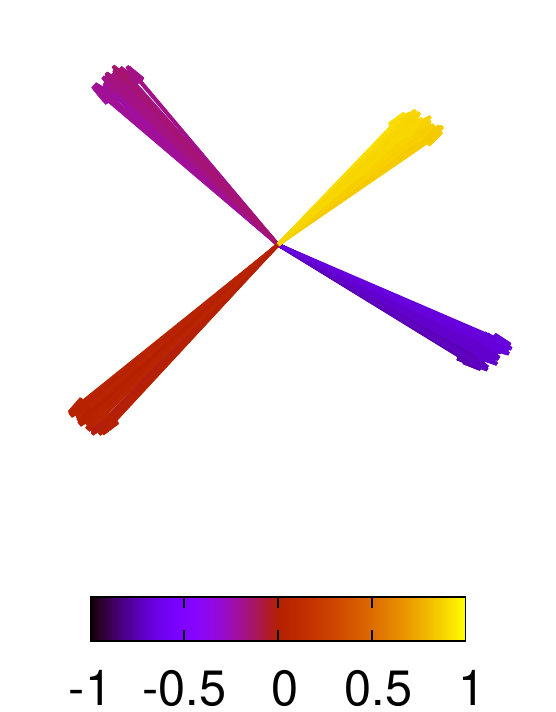}}\\
\caption{Chiral non-coplanar pattern at quarter filling for
  $t_{1}=-1$, $J_1=1.25 |t_1|$, and $J_2=0.5|t_1|$. (a) shows the spin
  pattern. The spin
  directions found fall along four non-coplanar directions, see
  (b). As can be seen in (a), all four directions arise on each
  sublattice, so that we find the pattern of Refs.~\onlinecite{Li:2012dt,PhysRevB.85.035414,PhysRevLett.108.227204}.
\label{fig:chiral_quarter_tNN_0}}
\end{figure}

\begin{figure}
\subfigure[]{\includegraphics[width=0.8\columnwidth,trim=0 0 0 0,clip]{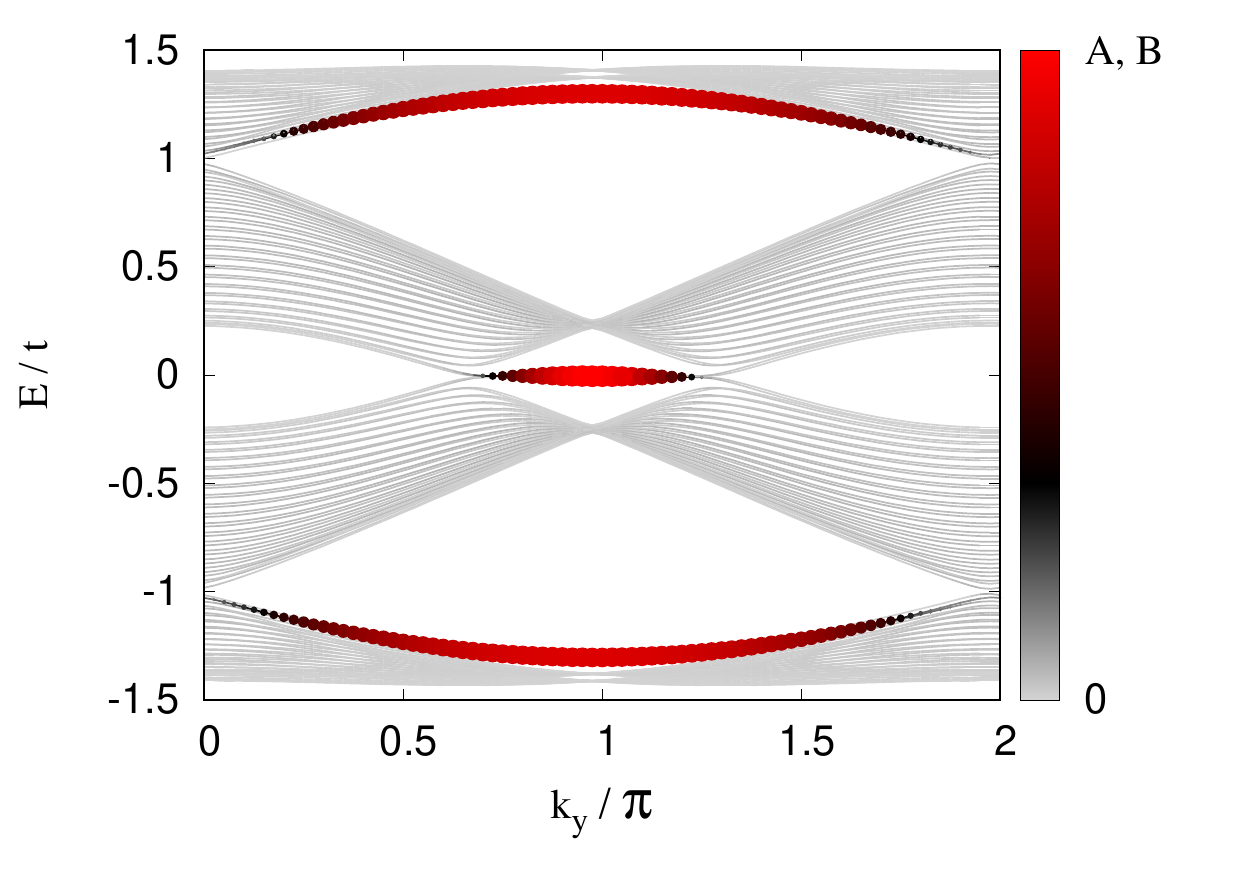}}
\subfigure[]{\includegraphics[width=0.8\columnwidth,trim=0 0 0 0,clip]{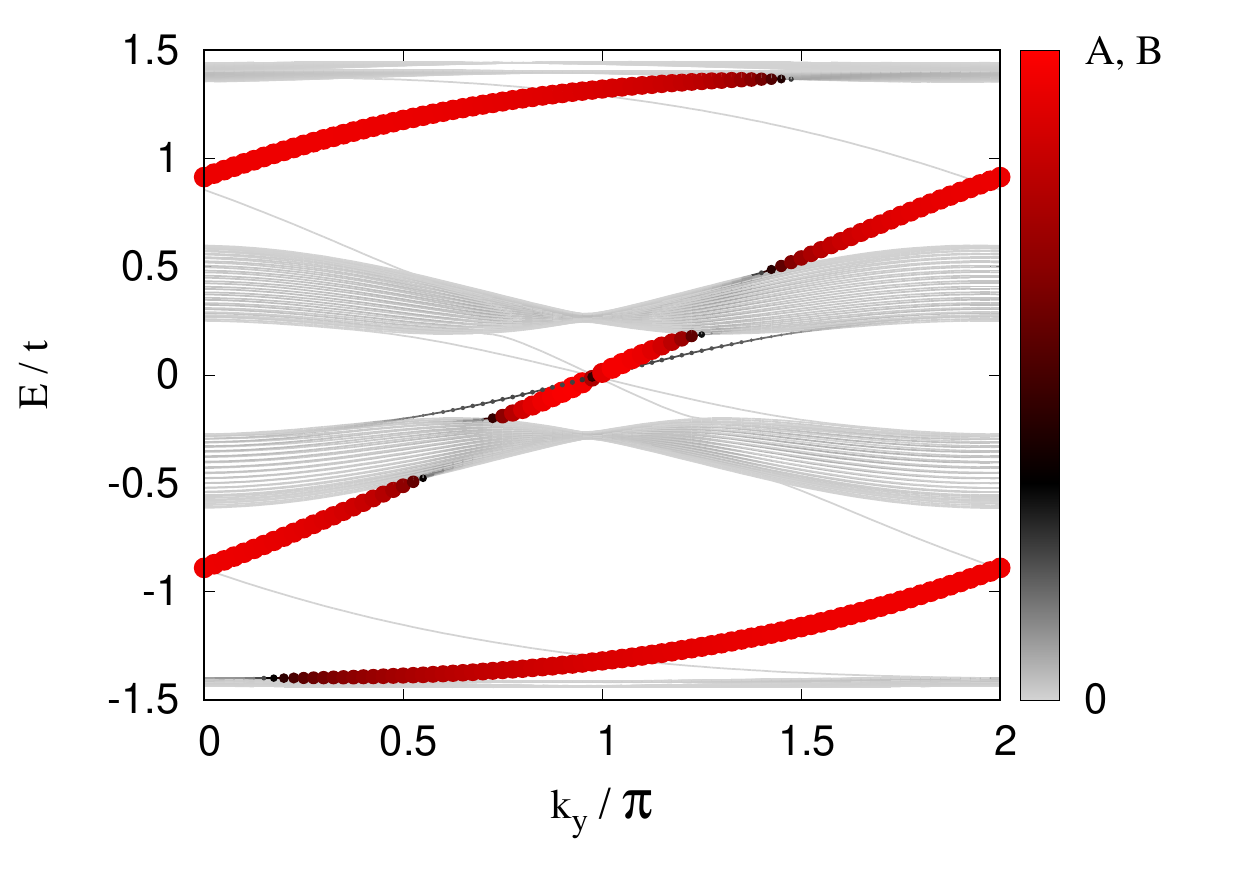}}
\caption{Edge states for the chiral pattern at quarter filling. Shown
  are eigenenergies of the phase shown in
  Fig.~\ref{fig:chiral_quarter_tNN_0} when put onto a cylinder. Width
  of the line indicates weight of the corresponding eigenstate on the
  top edge. In (a) $t_2=0$ and there are Dirac nodes, while
  $t_2=-0.2|t_1|$ opens gaps in (b).
\label{fig:edges_chiral_quarter}}
\end{figure}

At quarter filling, we observe the commensurate
non-coplanar pattern shown in Fig.~\ref{fig:chiral_quarter_tNN_0} if
two triangular Heisenberg models are coupled into a honeycomb system
by $t_1$ as well as  $J_1$. We find this phase to be
stable for $t_1 \lesssim J_1 \lesssim 3 t_1$ and $J_2 \approx 0.5 J_1$, i.e., for
roughly comparable interactions. It unit cell contains eight sites,
four on each sublattice. The four spins of the unit cell within one sublattice point
to the four corners of a tetrahedron and the same four directions
arise on both sublattices, see Fig.~\ref{fig:chiral_quarter_tNN_0}(b)
as third neighbors are always parallel. It has an inherent chirality
on each sublattice, and the two are parallel.

For the honeycomb \emph{Hubbard} model, the same pattern was discussed
at electron filling  $n=3/8$, where it is helped by the van-Hove
singularity,\cite{Li:2012dt,PhysRevB.85.035414,PhysRevLett.108.227204} and is expected to compete with a
collinear pattern at finite temperature.~\cite{PhysRevLett.108.227204}
In the Kondo-lattice model, the phase is not stabilized by a full gap, but by a
suppression of the density of states near the Fermi level:  four
subbands arise that are connected to their neighbors by Dirac
points. Correspondingly, fillings the density of states is lowered for
$n=1/4$ and $n=3/4$. (At $n=1/2$, nonmagnetic density of states
already has Dirac-cones and a pseudogap.) The van-Hove filling $n=3/8$
would, in contrast, lie in the middle of a subband, so that it does
here  not easily support such a state.

Allowing finite $t_2 \neq 0$ opens full gaps, see
Fig.~\ref{fig:edges_chiral_quarter}, while
still yielding the same spin pattern (not shown) in MCMC simulations. The gaps are topologically
nontrivial such that the system has now four bands, the lower two with  $C=1$
and the higher two with $C=-1$. (Or the other way around for inverted
chirality.) For a system put onto a cylinder, edge states
accordingly cross the bands gaps, two of them running in parallel for
the middle gap that would correspond to half filling. One 
edge runs across the Fermi level at quarter filling, and the system is
a Chern insulator with $|C|=1$. Coincidentally, $t_2=-0.2 t_1$ also
makes the lowest band very flat, see
Fig.~\ref{fig:edges_chiral_quarter}(b), yielding a promising stage to study
interacting and topologically nontrivial systems.~\cite{doi:10.1142/S021797921330017X} 

\subsubsection{Topologically nontrivial states from coupled triangular-lattice Chern insulators}

At half filling, the triangular
Kondo-lattice model with purely NN hopping $|t_1|=1$ supports a robust
QAH phase,~\cite{Martin:2008dx} for both AFM NN
exchange $J_1>0$~\cite{Kumar:2010p216405} and for finite Hund's rule
coupling.~\cite{Kato_FKLM_tri_2010} Its spin pattern corresponds to
one sublattice of the pattern discussed above, i.e., to one sublattice
of Fig.~\ref{fig:chiral_quarter_tNN_0}(a). The band gap at the Fermi
level that  stabilizes the
pattern separates two Chern bands with  $C= \pm 1$.

We now consider the case of coupling two such half-filled
triangular-lattice Chern insulators into a honeycomb system. Let
us for simplicity first discuss a  purely magnetic coupling, i.e. without hopping
between the sublattices. If the magnetic inter-lattice coupling is weak enough to preserve the chiral
non-coplanar pattern within each sublattice, we expect to find just two
copies of the Chern bands. The Chern numbers can either be the same
in both sublattices or opposite to each other. In the first case, the two edge states run
in parallel, in the second, they run opposite to each other: if the
state from sublattice $A$ runs to the right on the top edge, then the
state from sublattice $B$ runs to the left.

\begin{figure}
\subfigure[]{\includegraphics[width=0.6\columnwidth,trim=50 30 50 50,clip]{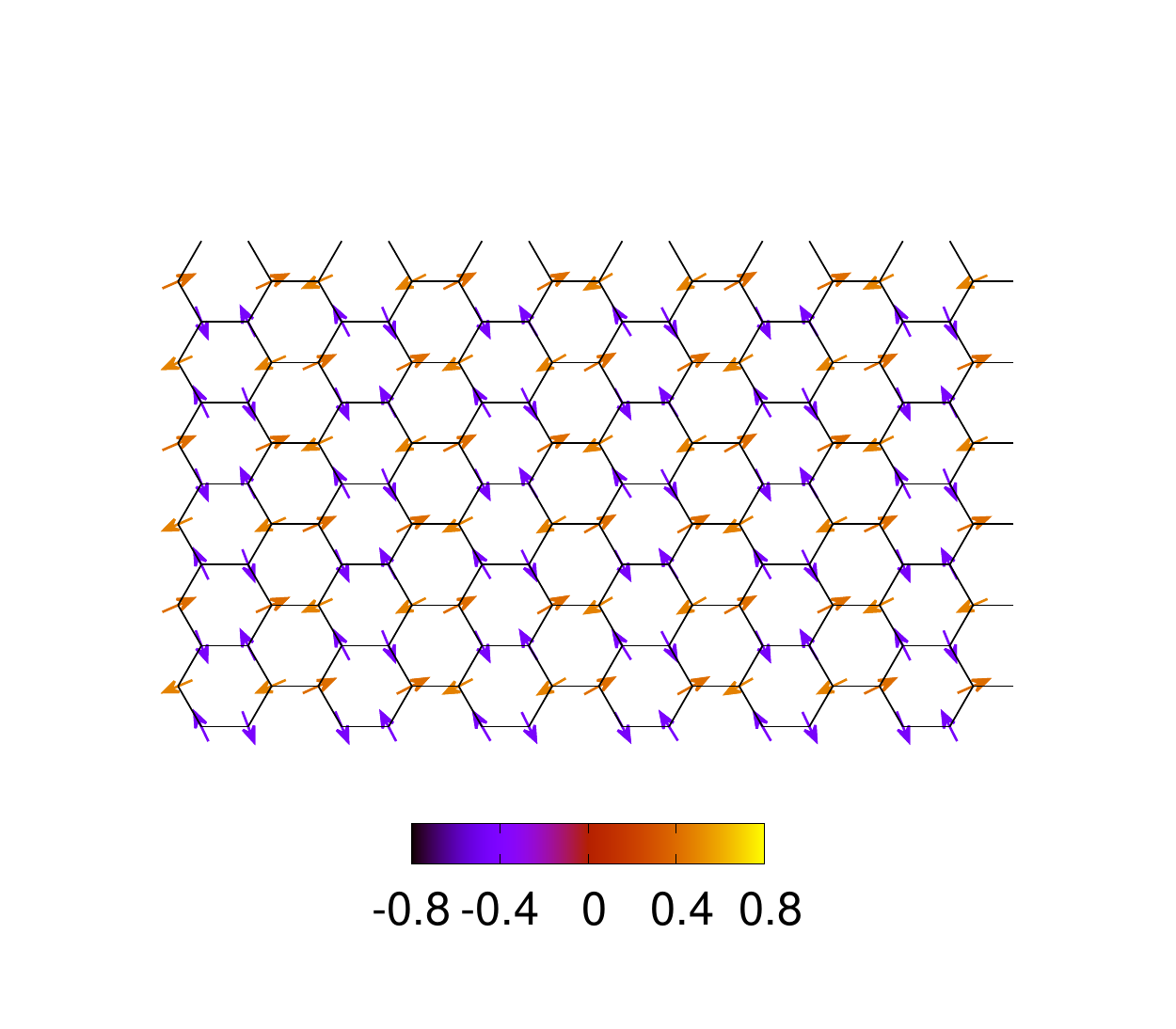}}
\subfigure[]{\includegraphics[width=0.37\columnwidth,trim=100 30 100 80,clip]{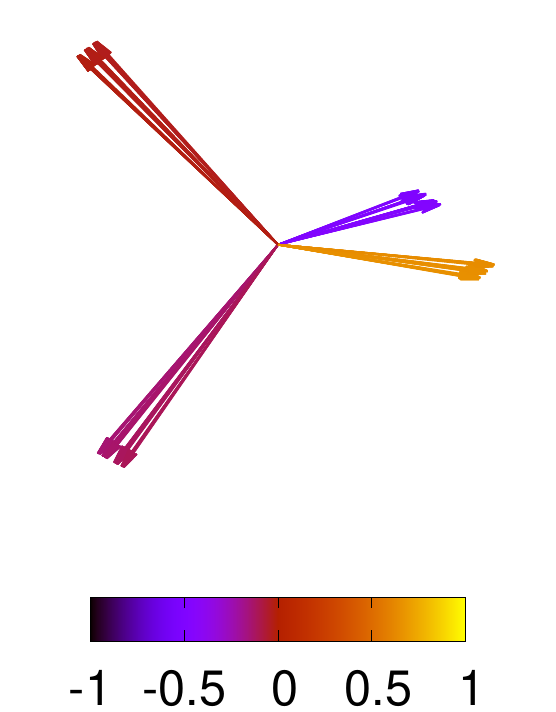}}\\
\subfigure[]{\includegraphics[width=0.8\columnwidth,trim=0 0 0 0,clip]{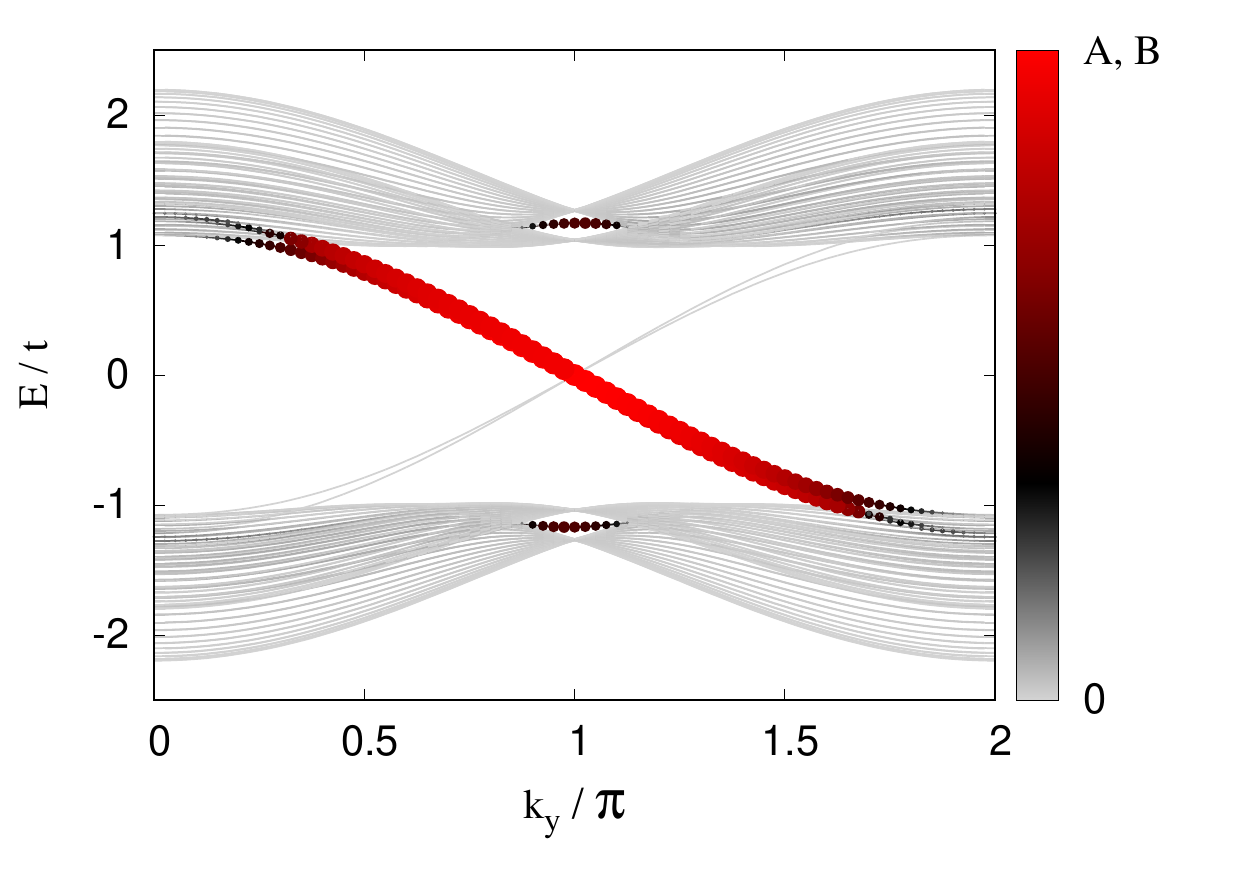}}
\caption{AFM coupled QAH states. Within each triangular sublattice,
  hopping $t_{2}=1$ and AFM Heisenberg exchange
  $J_2=0.15$ stabilize the non-coplanar QAH state. Coupling between
  sublattices is via both kinetic energy $t_{1}=0.2t_2$ and Heisenberg
  exchange $J_1=0.1$, resulting in
  an overall AFM coupling. As a result, (almost) the same four spin
  directions arise on both sublattices, see (b), scalar chiralities are the
  same and edge states on the two sublattices run in parallel. Width
  of the line in (c) refers to the weight of the left edge of the
  cylinder and illustrated that two of the edge states are well
  localized to this edge. (The other two run along the right edge.) \label{fig:chiral_tNN_02}}
\end{figure}

When the coupling between sublattices is via hopping, two questions
arise, (i) what effective magnetic coupling and consequently total
spin pattern arises and (ii) what is the topological character of the
combined band. Concerning the effective coupling, it
turns out to be weak, presumably due to the fact that each triangular
lattice is already gapped at the Fermi level. As 
the AFM and FM configurations turn out to be extremely close in
energy, with MCMC simulations occasionally fall in the wrong local
minimum. Direct energy comparison reveals that for infinite Hund's
rule coupling, the electron-mediated interaction is effectively FM, as one would expect for
a double-exchange mechanism, and grows with $|t_{1}|$. However, it is rather weak and other
exchange processes  could easily overcome it and induce
overall AFM coupling. A natural example would be  large
but finite Hund's-rule coupling $J_{\textrm{Hund}}$, where an 
effective AFM exchange $\propto t_{1}^2/J_{\textrm{Hund}}$ arises in second-order perturbation
theory~\cite{PhysRevB.66.144425}

As both AFM and FM coupling are thus potentially relevant, let us
discuss the overall spin pattern and band character in each case. 
First, we note that MCMC simulations, e.g. for $t_1=0.2 t_2$ and
$J_1=\pm 0.1$, indicate that the chiral spin pattern within the
sublattices is indeed preserved. Second,  AFM
and FM Heisenberg coupling between the sublattices selects identical
or opposite Chern numbers, resp.: For
each $B$-sublattice spin, the three $A$-sublattice spins occupying its
NN sites are along three of the four directions of the chiral
pattern; their combined magnetic moment points in the direction opposite
to the fourth and `missing' spin. For AFM $J_{1}>0$, the $B$ spin will
align itself opposite to this combined moment and consequently along said
fourth direction. As a result the same four directions arise in each
sublattice, see  Fig.~\ref{fig:chiral_tNN_02}(a,b), where third-neighbor
spins are always parallel. Chiralities and Chern numbers of the two
sublattices are then the same. Connection of two bands with $C=1$ into one band
with $C=2$ is natural, two edge state continue to run in
parallel,  see  Fig.~\ref{fig:chiral_tNN_02}(c).
A very similar phase
arising in a Hubbard-like model has been discussed in Ref.~\onlinecite{Jiang:2015dg}.

\begin{figure}
\subfigure{\includegraphics[width=0.6\columnwidth,clip]{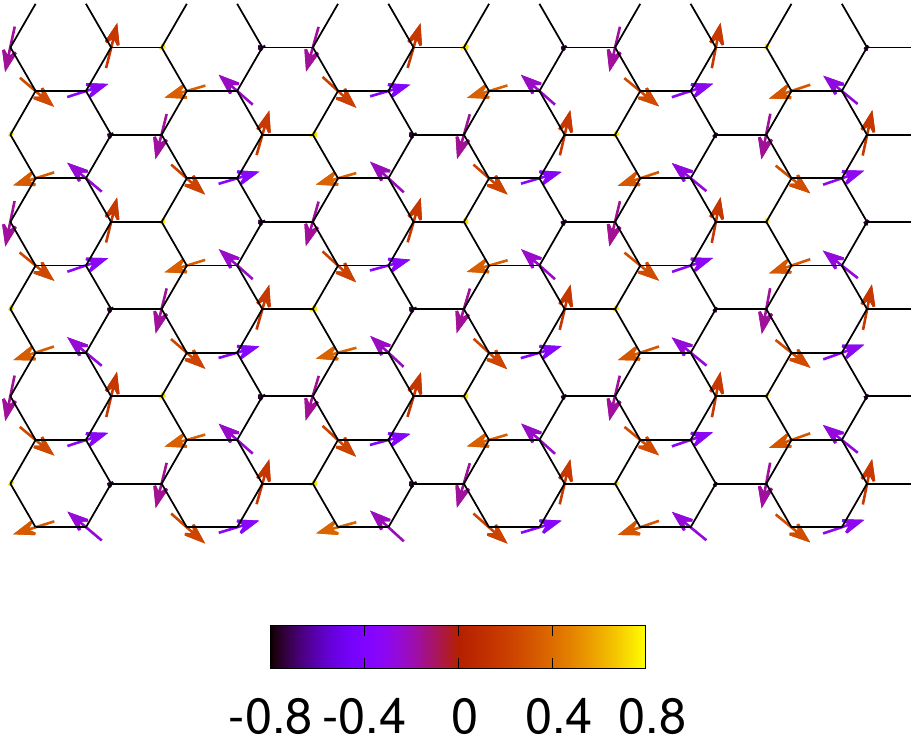}}
\subfigure{\includegraphics[width=0.37\columnwidth,trim=100 30 100 80,clip]{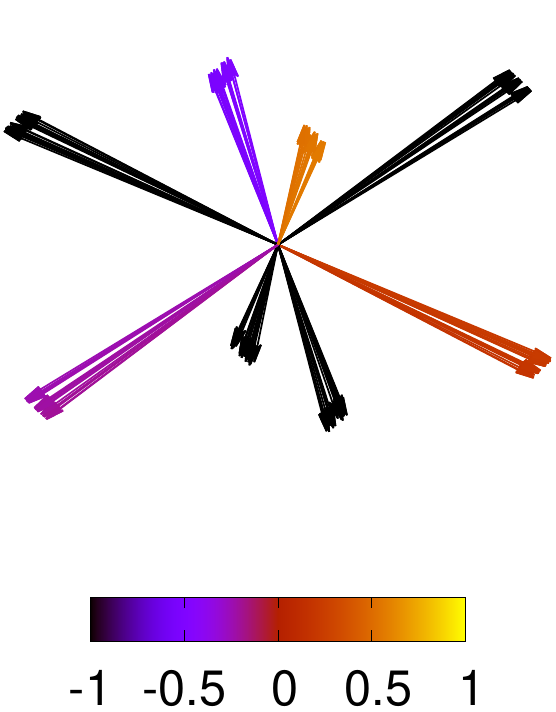}}\\
\subfigure{\includegraphics[width=0.8\columnwidth,trim=0 0 0 0,clip]{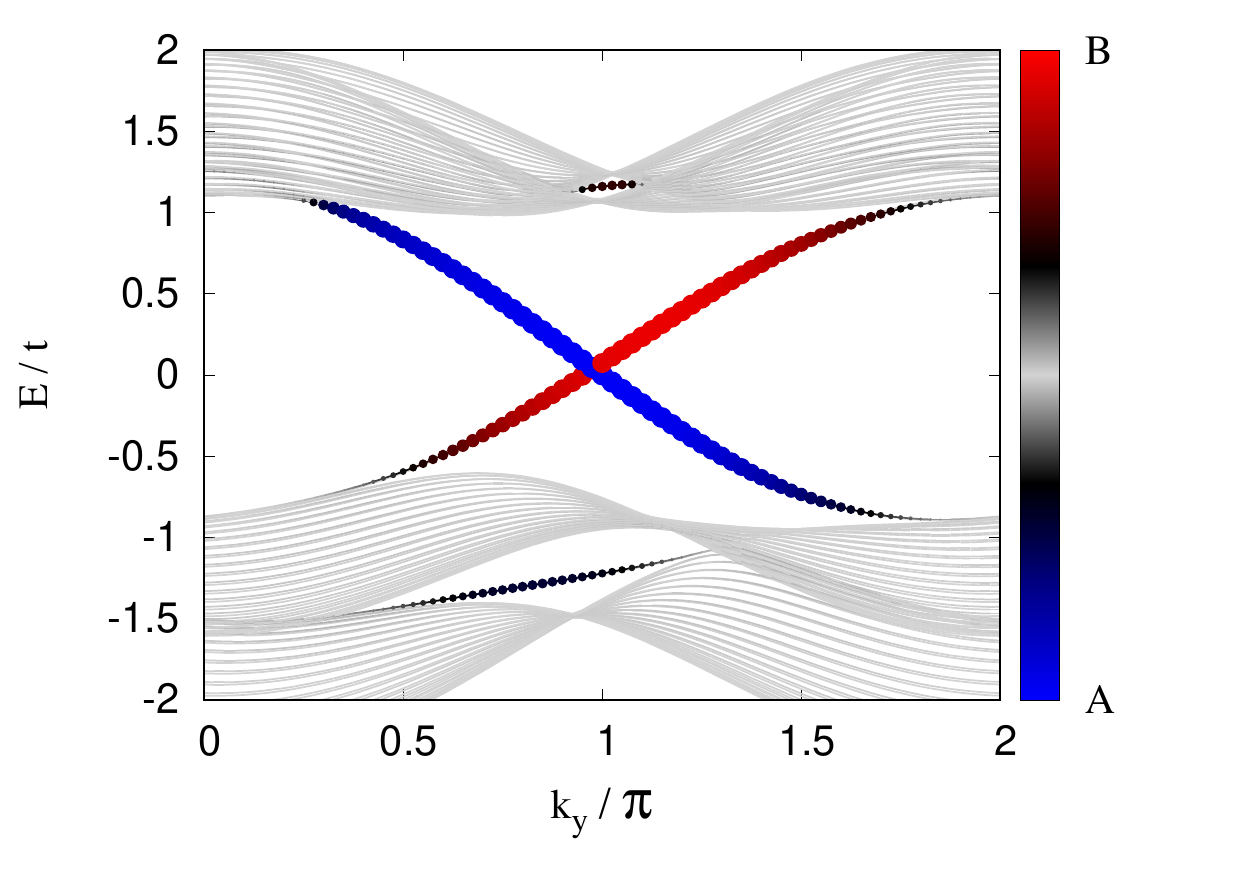}}
\subfigure{\includegraphics[width=0.8\columnwidth,trim=0 0 0 0,clip]{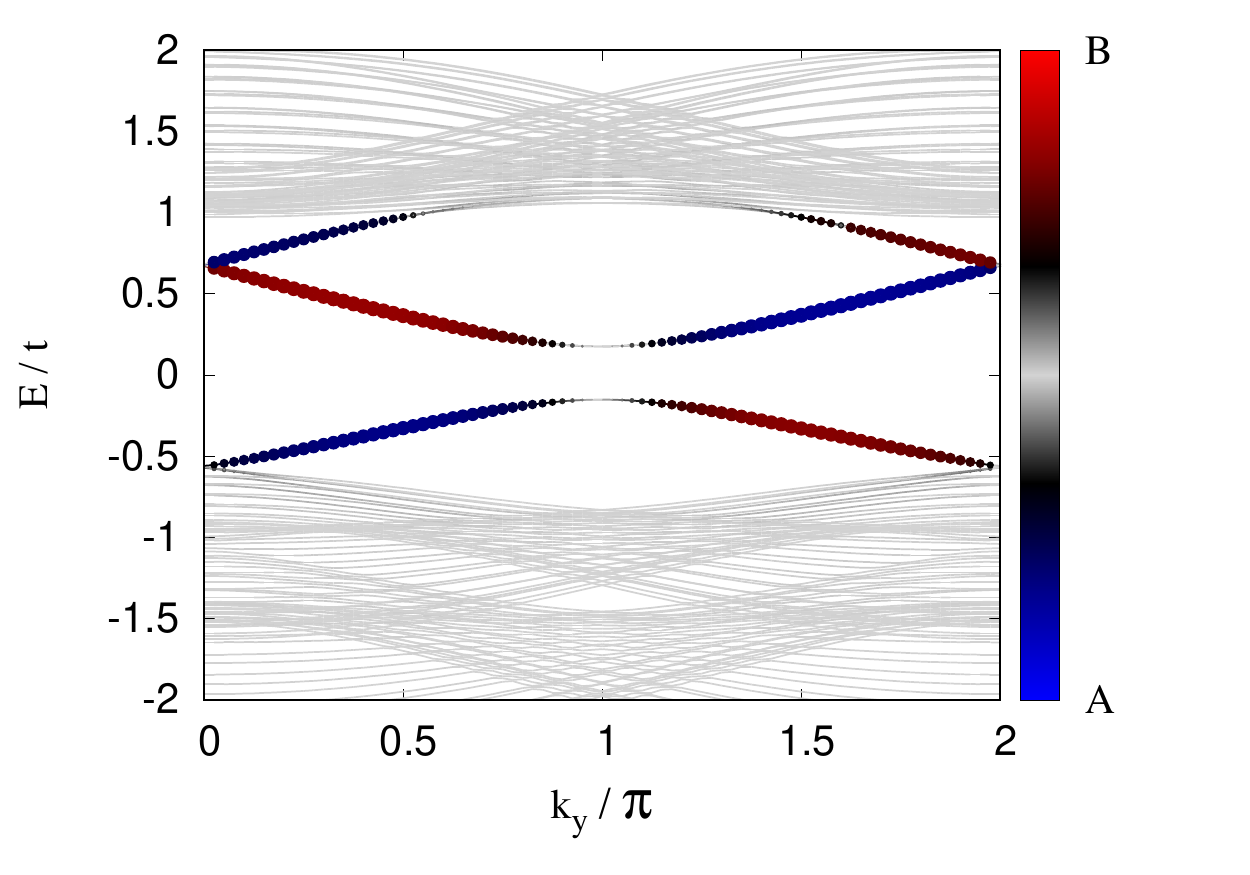}}
\caption{FM coupled QAH states  with $t_{1}= 0.2 t_2$ and $J_2=-0.1$,
  $t_2=1$, $J_2=0.15$. Spin orientations arising on the two
  sublattices in (a) are
  now antiparallel. In (b), where black arrows are for
  one of the sublattices and lighter ones for the other, this leads to a total of 8 different spin
  directions. Scalar chiralities are also opposite. Edge states along (c) zig-zag and (d)
  armchair edges. Line width indicates weight on the edge and shading is chosen according to sublattice character; the
  zigzag edge states are each located on one sublattice.  \label{fig:chiral_tNN_05}}
\end{figure}

\subsubsection{Antiferromagnetic topological crystalline insulator protected by
time-reversal plus reflection along $y$ direction}
\label{sec:TCI}

For FM
$J_{1}<0$, on the other hand, the $B$ spins will align
themselves  opposite to directions of the $A$ sublattice, making
all third-neighbor spins antiparallel. This gives  a
total of eight spin directions, see  Fig.~\ref{fig:chiral_tNN_05}(a,b), as
well as opposite chiralities. If there were no hopping between the
sublattices, i.e. for $t_1=0$, this would imply two bands with
opposite Chern numbers and edge states running in opposite
directions. Since NN hopping $t_1$ mixes the sublattices, a naturally resulting scenario is that the edge states
hybridize and  open a gap at the Fermi level when the two bands with $C=\pm 1$
are mixed  into a band with $C=0$. However, this is not
quite what happens, as can be seen in Fig.~\ref{fig:chiral_tNN_05}(c,d):
while the total Chern number is indeed $C=0$ and while armchair
edge states open a gap (as expected), zig-zag edge states do not
hybridize and  continue to cross each other as well as the Fermi
level. 

Such a behavior -- robust edge states on one edge, but not on the
other -- suggests a similarity to topological crystalline insulators:~\cite{Fu:2011ia}
As  a spatial symmetry takes the role that is played by time-reversal
invariance in topological insulators, edge states are only protected
on edges that share the relevant symmetry.  As we are going to see, the
phase is actually protected by a combination of a spatial reflection
and time inversion, similar to the case of a newly proposed
interacting topological state, the topological magnet.\cite{PhysRevB.95.081107}

Within each triangular sublattice, the effective hoppings
in the QAH phase show an internal symmetry, so that the four-site
magnetic unit cell can be reduced to a two-site electronic unit
cell.~\cite{Martin08} On the honeycomb lattice, the unit cell can thus
be reduced from eight to four sites. Labeling the two sublattices of the honeycomb lattice  by $+$ and
$-$ and the two sites of the electronic unit cell within each
triangular sublattice by $A$ and $B$, the kinetic energy
$\propto t_2$ can be written in terms of Pauli matrices $\tau^{x/y/z}$
acting on the electronic sublattice degree of freedom, i.e.,
\begin{align}\label{eq:tNNN_TCI}
  {H}_2=\sum_{\vec{k},\alpha=+,-}
\left(\cdag_{\vec{k},\alpha,A},  \cdag_{\vec{k},\alpha,B}\right)
\vec{d}_{\vec{k},\alpha}\cdot\vec{\tau}
\left(\begin{array}{c}
\cnod_{\vec{k},\alpha,A}\\ \cnod_{\vec{k},\alpha,B}
\end{array}\right)\;.
\end{align}
The vectors $\vec{d}_{\vec{k},\alpha}$ have entries
\begin{align}\label{eq:tNNN_TCI_2}
d_{\vec{k},\alpha}^z &= -\frac{2t_2}{\sqrt{3}}
  \cos(\vec{k}(\vec{a}_{1}-\vec{a}_{2}))\\
d_{\vec{k},\alpha}^x &= t_2\sqrt{\frac{2}{3}}
 (\cos \vec{k}\vec{a}_{1} + \cos \vec{k}\vec{a}_{2})\\
d_{\vec{k},\pm}^y &= \mp t_2\sqrt{\frac{2}{3}}
 (\cos \vec{k}\vec{a}_{1} - \cos \vec{k}\vec{a}_{2})\;,
\end{align}
where the sign of $d_{\vec{k},\alpha}^y$ chooses one of the two chiralities
and fixes the Chern number of the occupied band. Time reversal acts on
this sign. Due to a local gauge freedom, other parameterizations~\cite{Kourtis:2012hn} are
possible, but the present one leads to a more symmetric NN hopping, see
below. The fact that honeycomb sublattices $+$ and $-$ do not mix,
gives Eq.~(\ref{eq:tNNN_TCI}) an accidental time-reversal symmetry
that is removed by $t_1\neq 0$.

The kinetic energy of the NN interlattice hopping can be written as
\begin{align}\label{eq:tNN_TCI}
  {H}_1=\sum_{\vec{k}}
\left(\cdag_{\vec{k},+,A},  \cdag_{\vec{k},+,B}\right)
\left(\begin{array}{cc}
h^{+-}_{AA}& h^{+-}_{AB}\\
h^{+-}_{BA} & h^{+-}_{BB}
\end{array}\right)
\left(\begin{array}{c}
\cnod_{\vec{k},-,A}\\ \cnod_{\vec{k},-,B}
\end{array}\right) + \textrm{H.c}\;
\end{align}
with matrix elements 
 \begin{align}
 h^{+-}_{AA}(\vec{k}) &= -i(\e^{-i\frac{\pi}{12}}\e^{i\vec{k}\vec{b}_1}-\e^{i\frac{\pi}{12}}\e^{i\vec{k}\vec{b}_2})\\
 h^{+-}_{BB}(\vec{k}) &= -\e^{i\frac{\pi}{12}}\e^{i\vec{k}\vec{b}_1}+\e^{-i\frac{\pi}{12}}\e^{i\vec{k}\vec{b}_2}\\
 h^{+-}_{AB}(\vec{k}) &= h^{+-}_{BA}(\vec{k}) = -\e^{-i\vec{k}(\vec{b}_1+\vec{b}_2)}\;.\label{eq:end_eff_ham}
 \end{align}
Time reversal inverts all spins and thus gives the complex conjugate of the position-space hopping
elements, but does not take the complex conjugate of the Fourier
factors $\e^{i\vec{k}\vec{b_i}}$ that are purely due to spatial
geometry. The time-reversed matrix elements become thus
 \begin{align}
\mathcal{T} h^{+-}_{AA}(\vec{k}) &= i(\e^{i\frac{\pi}{12}}\e^{i\vec{k}\vec{b}_1}-\e^{-i\frac{\pi}{12}}\e^{i\vec{k}\vec{b}_2})\\
\mathcal{T} h^{+-}_{BB}(\vec{k}) &= -\e^{-i\frac{\pi}{12}}\e^{i\vec{k}\vec{b}_1}+\e^{i\frac{\pi}{12}}\e^{i\vec{k}\vec{b}_2}\\
\mathcal{T} h^{+-}_{AB}(\vec{k}) &=\mathcal{T} h^{+-}_{BA}(\vec{k}) = -\e^{-i\vec{k}(\vec{b}_1+\vec{b}_2)}=h^{+-}_{AB}(\vec{k})\;.
 \end{align}

\begin{figure}
\includegraphics[width=0.7\columnwidth,clip]{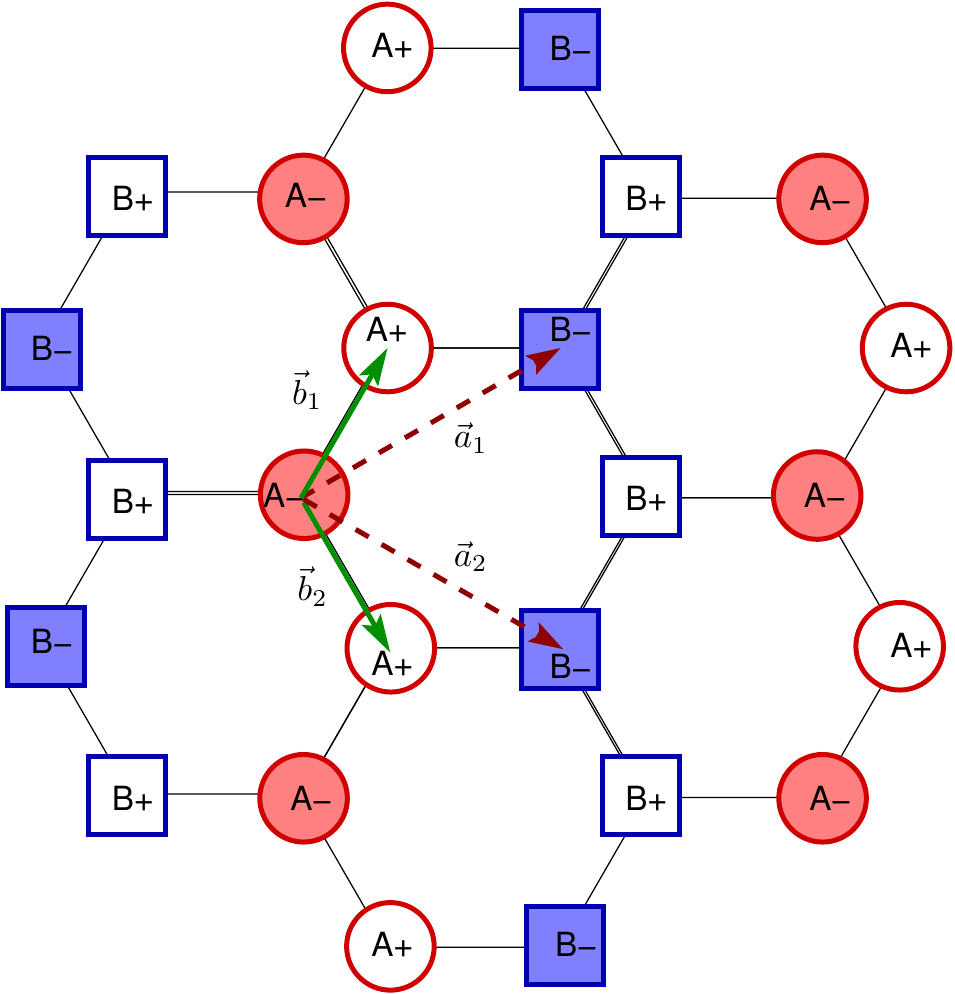}
\caption{Sublattice structure underlying the effective electronic
  Hamiltonian Eqs.~(\ref{eq:tNNN_TCI}) - (\ref{eq:end_eff_ham}). Empty and filled symbols (resp. $+$ and $-$, as in the
  main text) indicate the triangular sublattices of the
  honeycomb lattice, $\vec{a}_1$ and $\vec{a}_2$ are basis vectors of
  the underlying Bravais lattice,
  $\vec{b}_1=(2\vec{a}_1-\vec{a}_2)/3$ and
  $\vec{b}_2=(2\vec{a}_2-\vec{a}_1)/3$ connect NN sites. Circles and
  squares indicate the magnetic $A$ and $B$ sublattices within each
  triangular sublattice. Inversion of the $y$ direction switches
  $\vec{b}_1$ with $\vec{b}_2$ and $\vec{a}_1$ with $\vec{a}_2$.\label{fig:geometry_TCI}}
\end{figure}

NN hopping Eq.~(\ref{eq:tNN_TCI}) is clearly not time-reversal
invariant as $\mathcal{T} h^{+-}_{XX}(\vec{k})\neq h^{+-}_{XX}(\vec{k})$, but a simultaneous exchange $\vec{b}_1\leftrightarrow \vec{b}_2$ restores the original matrix
elements. As can be seen in Fig.~\ref{fig:geometry_TCI}, the
corresponding operation is a reversal of $y$ direction, i.e., of the
direction along a zig-zag edge. This reflection exchanges at the same
time vectors $\vec{a}_1\leftrightarrow \vec{a}_2$ of the triangular
sublattices. Equations~(\ref{eq:tNNN_TCI}) and~(\ref{eq:tNNN_TCI_2})
indicate that this operation changes the sign of $d_{\vec{k},\pm}^y$
and thus likewise undoes the effect of time reversal. Both parts of
the Hamiltonian are accordingly invariant under a combination of time
reversal and reflection of the $y$-direction and this symmetry
protects the edge states along zig-zag edges. Both time and $y$ reversal invert the sign
of component $k_y$ along the zig-zag edge, so that the total combined
symmetry operation keeps the sign of $k_y$ intact. As a result, the
crossing of the edge states is not pinned to high-symmetry momenta
like $k_y=0,\pi$, but can -- and does -- occur at arbitrary momenta
depending on $t_1$, see Fig.~\ref{fig:chiral_tNN_05}.

Naturally, the question of invariants arises. The total Chern number of these two bands is 0, as discussed
above, and in the presence of a sublattice-mixing $t_1\neq 0$, there is
also no sublattice Chern number to take its place.  While it is not
apparent from Fig.~\ref{fig:chiral_tNN_05}, the two occupied bands are
almost separated from one another, connected only by a Dirac
point. A `mass term', i.e. a staggered onsite potential of
opposite sign for the two sublattices, opens a gap at this Dirac
point, so that Chern numbers for the two occupied bands can readily be
obtained numerically. They are opposite, i.e., one of the occupied (as
well as of the empty) bands has $C=1$ and the other $C=-1$. 

\begin{figure}
\subfigure{\includegraphics[width=0.5\columnwidth]{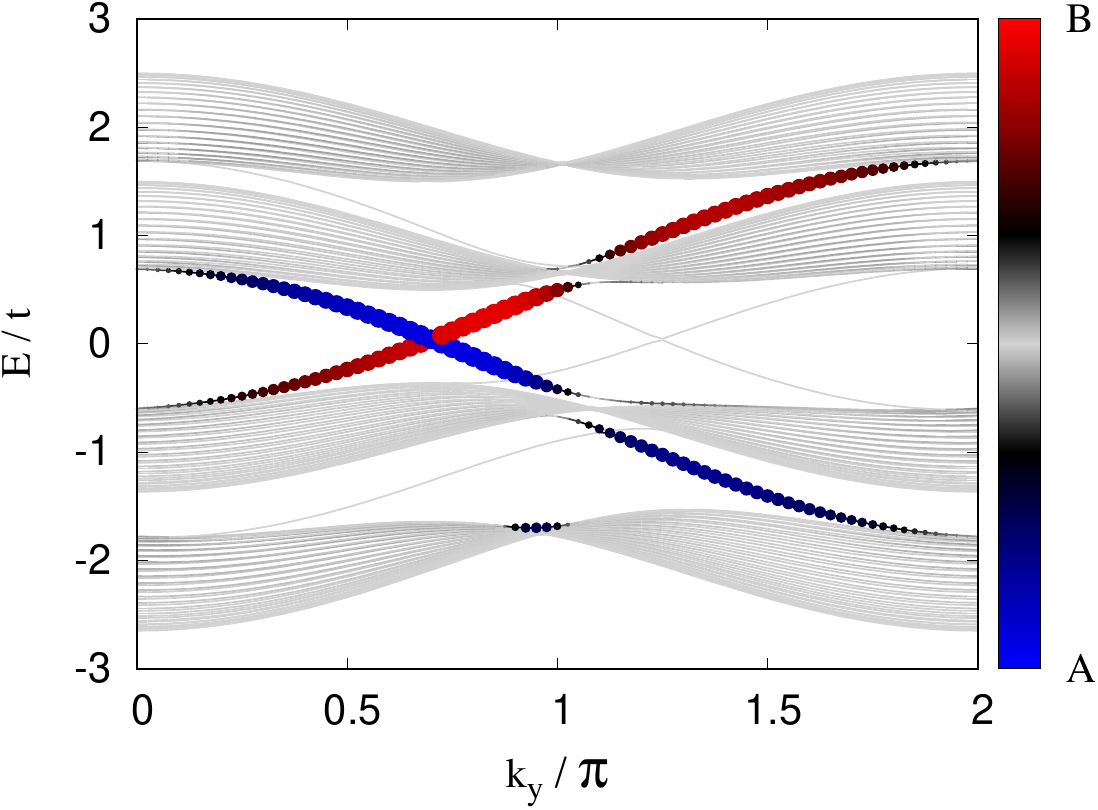}}
\subfigure{\includegraphics[width=0.5\columnwidth]{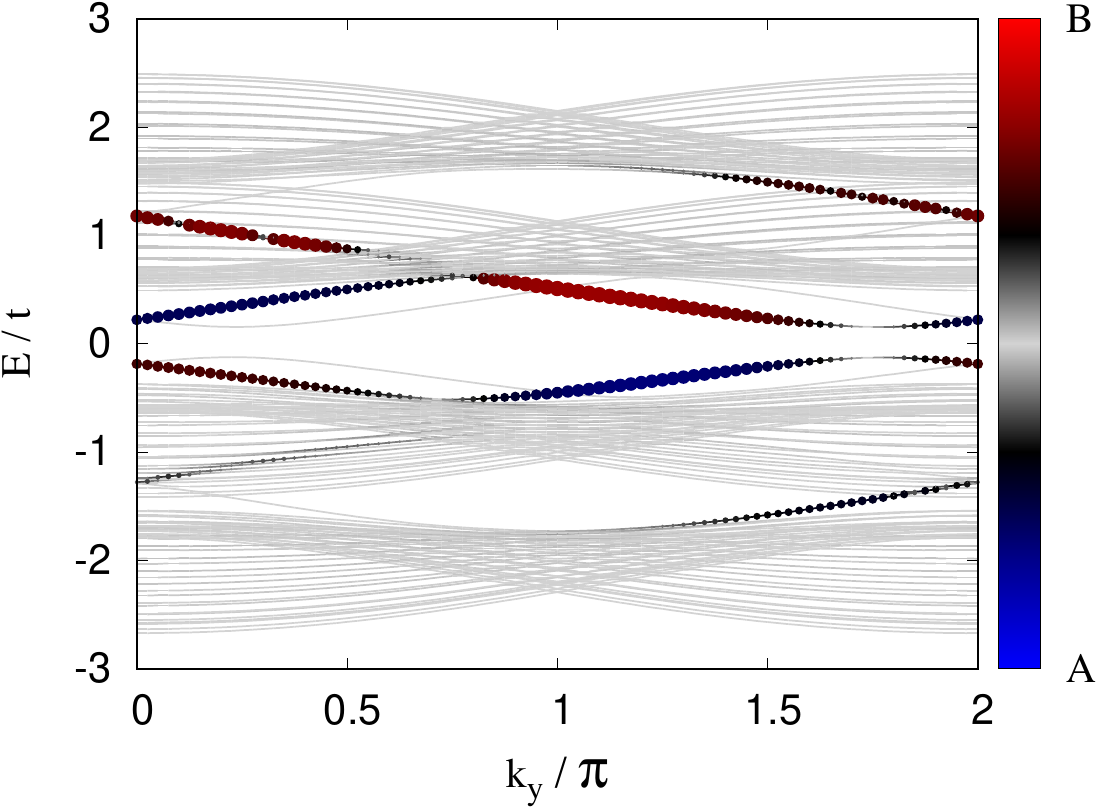}}
\caption{Edge states for (a) zigzag and (b) armchair edges and the
  pattern with FM coupling and opposite chirality between the $+$ and
  $-$ sublattices. Parameters $t_{1}= 0.2 t_2$, $J_1=-0.1$, $J_2=0.15$,
  $t_2=1$ as in 
  Fig.~\ref{fig:chiral_tNN_05}(c,d), but with an additional staggered
  onsite potential $m=0.5 t_2$ that makes $+$ and $-$ sublattices
  inequivalent. Monte-Carlo results thus indicate that the phase is
  stable for moderate $m\neq 0$.}
\end{figure}

Figure~\ref{fig:chiral_tNN_05} shows the edge states for such a case,
here with a substantial staggered onsite potential. Making the honeycomb
sublattices $+$ and $-$ inequivalent clearly does not affect the
topological protection, as also expected from the above symmetry
considerations. The zigzag edge states connect pairs of empty and
occupied subbands with opposite Chern numbers. Both with and without
onsite potential, the edge states are localized on one of the
sublattices, even though the bulk bands, which they connect, mix sublattices.  This is
reminiscent of graphene (without spin-orbit coupling), where the edge-states on zigzag and bearded
edges  can likewise be assigned to a sublattice and persist for inequivalent
sublattices.~\cite{Yao:2009p2427} In the graphene case, the edge states do not generically
cross the Fermi level. Here, where the additional structure ($A$
vs. $B$) within each sublattice drives the band gap at the Fermi
level, they do.

For simplicity, we
have here restricted the discussion to the case of infinite Hund's-rule coupling,
where only the spin projection parallel to the local magnetic order is
active. If Hund's-rule coupling is large, but finite, the hoppings
between parallel and antiparallel spins require us to use the full
eight-site magnetic unit cell in addition to bringing both spin
directions into play. The $16\times 16$ matrix makes the discussion more
cumbersome, but we have verified that  the physics reported here
(e.g. protected edge states on zigzag edges) remains valid.

\section{Summary and Conclusions} \label{sec:conclusion}

Motivated by materials where frustrated spin and electron systems can
be coupled into an unfrustrated, or at least less frustrated, total
system, we study the double-exchange model with NN and NNN hopping and
spin exchange on the honeycomb lattice. We focus first on the interplay of NN
hopping (promoting FM order) with NNN superexchange (frustrated, promoting AFM
with a $120^\circ$ Yaffet-Kittel pattern): For half filling, the phase diagram
is dominated by coplanar, but non-collinear, patterns mostly built up from AFM dimers. For all theses
states as well as for the limiting FM and YK AFM states, the effective
electronic bands remain graphene-like with Dirac cones at the Fermi
level. 

When discussing a filling of $n=1/3$ or $n=2/3$ rather than half
filling, density of states at the
Fermi level of the FM state is quite high  and we would thus
expect the kinetic energy to have a larger impact than at half filling. To
some extent, this is indeed what happens: Finite $J_2$ here induces non-coplanar spin
patterns. Such a  tendency to non-coplanar order has been observed in frustrated
double-exchange models before and is an important difference to purely magnetic frustrated Heisenberg
honeycomb models, where states remain coplanar. On the other hand, we find
here mostly incommensurate patterns, in contrast to $n=1/2$ and similar to
Heisenberg models. 

We then turned our investigation to coupling two triangular-lattice quantum
anomalous-Hall states into a honeycomb lattice. For AFM coupling, a
Chern insulator with $C=2$ arises, while FM coupling gives $C=0$. We find that NN
hopping mediates only very small FM coupling. The resulting phase may be related to the  $C=0$ state with non-coplanar magnetism
recently reported for the honeycomb Hubbard model~\cite{Jiang:2015dg}
and is, however, not topologically trivial despite its vanishing Chern
number, see Sec.~\ref{sec:TCI}. Instead, a combination of time-reversal and mirror-reflection
symmetries protects edge states on zig-zag, but not armchair,
edges.

The combination of time-reversal and inversion/reflection symmetry
protecting a topologically nontrivial state has
recently also been discussed as crucial for topologically nontrivial phases in
double-perovskites bilayers.~\cite{Dong:2016vd} The scenario differs
in several instances from our, e.g. concerning the character of the
state in question and the origin of nontrivial band
topology. Nevertheless, it is intriguing to note that both are 
systems where frustrated sublattices are connected and where this
interplay leads to new topological phases driven by crystal symmetry
together with (some variant of) antiferromagnetism.

\begin{acknowledgments}
We gratefully acknowledge stimulating and helpful discussion with J. Venderbos and
J. van den Brink. This research was supported by the Deutsche
Forschungsgemeinschaft,  
via the Emmy-Noether program (DA 1235/1-1) and FOR1807 (DA 1235/5-1);
by the NSF through Grant Nos. DMR-1506263 and DMR-1506460. SR acknowledges ITF, IFW for computing cluster.
\end{acknowledgments}


%

\end{document}